\documentclass[twocolumn]{aastex62}
\usepackage[utf8]{inputenc}

\usepackage[nointegrals]{wasysym}
\usepackage{amsmath}
\usepackage{graphicx}
\usepackage{commath}
\usepackage{bm}
\usepackage{float}
\usepackage[caption=false]{subfig}

\usepackage{savesym}
\savesymbol{tablenum}
\usepackage{siunitx}
\restoresymbol{SIX}{tablenum}

\begin{document}

\title{MULTI-EPOCH ULTRAVIOLET HST OBSERVATIONS OF ACCRETING LOW-MASS STARS}
\author[0000-0003-1639-510X]{Connor E. Robinson}
\author[0000-0001-9227-5949]{Catherine C. Espaillat}
\affiliation{Boston University, 725 Commonwealth Avenue, Boston, MA 02215, USA}

\shortauthors{Robinson \& Espaillat}
\shorttitle{HST OBSERVATIONS OF ACCRETING LOW-MASS STARS}
\email{connorr@bu.edu}

\begin{abstract}
Variability is a defining characteristic of young low-mass stars that are still accreting material from their primordial protoplanetary disk. Here we present the largest \textit{HST} variability study of Classical T Tauri stars (CTTS) to date. For 5 of these objects, we obtained a total of 25 spectra with the Space Telescope Imaging Spectrograph (STIS). Mass accretion rates and the fraction of the star covered by accretion columns (i.e., filling factors) were inferred using 1D NLTE physical models whose parameters were fit within a Bayesian framework. 
On week long timescales, typical changes in the mass accretion rates range up to a factor of $\sim2$, while changes of up to a factor of $\sim5$ are inferred for the filling factors.
In addition to this, we observed a possible accretion burst in the transitional disk system GM Aur, and an incident we interpret as a chance alignment of an accretion column and the undisturbed photosphere along our line of sight in the full disk system VW Cha. We also measure correlations between mass accretion rate and line luminosities for use as secondary tracers of accretion. We place our objects in context with recent high-cadence photometric surveys of low-mass star formation regions and highlight the need for more broad-wavelength, contemporaneous data to better understand the physical mechanisms behind accretion variability in CTTS.

\end{abstract}
\section{INTRODUCTION}

Classical T Tauri Stars (CTTS) are young low-mass stars still accreting material from their primordial protoplanetary disk \citep[see ][for a review on accretion in young stars]{hartmann16}. 
Some of the variability seen in CTTS may be due to a change in the mass accretion rate \citep[e.g.,][]{herbst94, cody14}. 
However, The driving forces for variable mass accretion rates and the effect of a changing high energy radiation field on the disk are currently not fully understood. 

Accretion onto CTTS is thought to occur through the mechanism of magnetospheric accretion \citep{shu94,koenigl91}.
Young stars are known to host magnetic fields with strengths on the order of several kG which are capable of disrupting the inner gas disk at a distance of a few stellar radii \citep{johnskrull00, donati10}.
The material bound to the magnetic field line is funneled from the inner regions of the disk towards the magnetic footprint on the star. As material falls along the field line towards the star, it reaches supersonic free-fall speeds ($\sim 300 \, km/s$) and forms a standing shock where it collides with the surface of the star. 

The accretion shock region transforms the kinetic energy of the column into an observed continuum excess through the following processes.
The kinetic energy of the column is transformed into thermal energy in the shock, heating the gas in the post-shock region to temperatures of $\sim10^6K$. As the gas falls and cools, the energy is primarily re-emitted as X-rays both towards and away from the star  \citep{calvet98}. The X-ray emission traveling toward the star heats the underlying photosphere, forming a hot spot. Emission traveling away from the star heats the material in the pre-shock region to temperatures of $\sim 10^4K$. 
These regions re-emit the energy of the accretion column into the observed far ultraviolet (FUV), near ultraviolet (NUV), and optical continuum excess \citep[for a review, see][]{hartmann16}. This, along with stronger emission lines from accretion processes, distinguishes CTTS from non-accreting Weak-Lined T Tauri Stars (WTTS).

While the bolometric luminosity of the UV excess directly depends on the amount of material accreted by the star, the spectral shape of this excess is a function of the density of the accretion column \citep{calvet98}. Higher density columns emit more FUV/NUV emission while lower density columns more closely resemble undisturbed photospheric emission, making UV spectra a useful tool for estimating density and surface coverage of accretion columns. These qualities in turn inform us on conditions in the inner regions of the disk. Previous studies have found it necessary to include contributions from 
from accretion columns with a range of energy fluxes in order to explain the observed UV excess \citep{ingleby13, ingleby15}.

Young low-mass stars often exhibit a rich forest of atomic and molecular FUV/NUV emission features. Although present, these lines are not nearly as strong in non-accreting Weak-Lined T Tauri Stars (WTTS), indicating that these lines are primarily produced through the act of accretion \citep{johnskrull00}.
$C_{IV} \, \lambda 1549\si{\angstrom}$, $He_{II} \, \lambda 1640\si{\angstrom},$ and $H_\alpha \, \lambda 6563\si{\angstrom}$ are several examples of features that have been shown to correlate with mass accretion rate \citep{johnskrull00, calvet04, ingleby11, yang12, ardila13, ingleby13}. These lines are useful secondary predictors of mass accretion rate when spectra with complete UV to optical coverage are unavailable. In addition, these lines provide information about the kinematics of the accretion shock \citep{ardila13}.

Several morphological classifications of variability in CTTS have been identified by photometric monitoring campaigns \citep[e.g.,][]{herbst94, alencar10, cody14}. The configuration of the magnetic field is likely in part responsible for the density and degree of surface coverage of the accretion columns \citep{adams12}. Spectropolarimetric measurements of accreting young stars indicate a variety of magnetic field configurations, ranging from well-ordered dipolar fields to complex non-axisymmetric field configurations \citep[e.g.,][]{donati10, donati11}.
Periodic accretion signatures associated with well-ordered magnetic fields are also present among objects within these surveys.
The collimation by this field is thought to result in accretion columns that impact the star at high stellar latitudes, giving rise to approximately sinusoidal light curves at the rotational frequency \citep{kulkarni08, romanova08}.
However, many other CTTS exhibit more chaotic signatures with limited or no obvious periodicity with characteristic timescales of a few days. 
Two classes of objects with more chaotic light curves that have been identified in previous photometric surveys are ``bursters" and ``stochastic objects" \citep[e.g.,][]{cody14,cody18}. 
These objects exhibit rapid changes in brightness which have been attributed to changes in the mass accretion rate. Stochastic objects tend to show both fading and brightening events, while bursters primarily exhibit brightening events. The median timescales for events from periodogram analysis for both of these types of objects is  $\sim20$ days, but individual sources have been identified with characteristic timescales ranging from $~\sim2 - 80$ days. These sorts of timescales make it possible to capture interesting large-scale changes in the accretion rate using observations with roughly week-long cadences. It is also worth noting that smaller scale accretion variability is also very common in young stars, with timescales ranging down to just a few minutes \citep[e.g.,][]{siwak18} which is missed by observations with longer cadences and integration times.

Simulations of magnetospheric accretion onto young stars predict various sources of accretion variability from a variety of instabilities in the inner regions of disks and the magnetosphere \citep{kulkarni08, romanova12, robinson17}. Inferences about the magnetic field structure and sources of variability can be made with measurements of the surface coverage of accretion columns from UV spectra of accreting objects while simultaneously measuring the mass accretion rate.

Previous studies with simultaneous FUV to NIR coverage have presented samples of single observations of several objects \citep{ingleby13, ardila13, Thanathibodee18}, or have interpreted repeated observations of a single object \citep{ingleby15}.
Multiple observations of individual objects mitigate systematic effects (e.g., distance and extinction), are required for measurements of variability, and place irregular accretion epochs in context when they might otherwise be considered standard behavior. On the other hand, broader surveys that include observations of many objects provide context for identifying trends and outliers in the global population of young stars \citep[e.g., identifying trends with accretion with age;][]{hartmann98}. In short, multi-epoch studies with multiple objects are necessary to fully interpret accretion variability within a local and global context. 

Here we present and model contemporaneous FUV-NIR observations of a sample of five well-studied CTTS. In \S 2 we present our sample and describe the observations. In \S 3 we briefly describe the \citet{calvet98} accretion shock model and discuss our updates to that model. We describe our fitting methods and present mass accretion rates, surface coverages and line luminosities in \S 4. Interpretation of our results is presented in \S 5, and we conclude in \S 6. 

\section{SAMPLE AND OBSERVATIONS}

\subsection{Objects}
The sample contains five young accreting stars surrounded by full, transitional (disks with an inner hole), and pre-transitional disks \citep[disks with gaps e.g.,][]{espaillat14}. The objects in this sample include DM Tau, GM Aur, SZ 45, TW Hya, and VW Cha.  These objects are all relatively well studied and associated with nearby star formation regions. More discussion on each object is presented in \S 5.2.

\subsection{Weak-Line T Tauri Stars}
WTTS are used as spectral templates in this analysis because even non-accreting young stars can have significant chromospheric emission in the UV above main sequence dwarf stars \citep{ingleby11}. Here we use spectra from three WTTS with spectral types similar to those of our CTTS targets: RECX 1 (K5), HBC 427 (K7), and TWA 7 (M1). All of these objects have V-band extinctions, $A_V$, close to 0 and have previously been used as spectral standards within the framework of the \citet{calvet98} shock models \citep{ingleby13}. These spectra were obtained as part of a large \textit{Hubble Space Telescope} (\textit{HST}) program: Disks, Accretion, and Outflows of T Tau stars (PI: G. Herczeg, proposal ID: 11616).

\subsection{HST STIS}
For each object, we obtained multi-epoch low resolution spectra covering roughly $1100\si{\angstrom}$ to $10000\si{\angstrom}$ using the Space Telescope Imaging Spectrograph (STIS) aboard the \textit{HST}. The spectra were primarily taken using 4 overlapping grating settings: FUV-MAMA G140L, NUV-MAMA G230L, CCD G430L and CCD G750L, which cover the FUV, NUV, optical and NIR, respectively. The observations for each grating were obtained in consecutive orbits, resulting in a near-contemporaneous dataset for each observational epoch. 
We note that these observations are not truly simultaneous since they are taken over several \textit{HST} orbits spanning roughly 2.5 hours and photometric and spectral variability on shorter timescales has been observed in CTTS \citep[e.g.,][]{bouvier03, alencar05, cody14, dupree12, siwak18}. In regions where gratings overlap we find good agreement between orbits, suggesting that this effect is minimal within this set of observations.

\begin{figure}
    \centering
    \includegraphics[width = 1\linewidth]{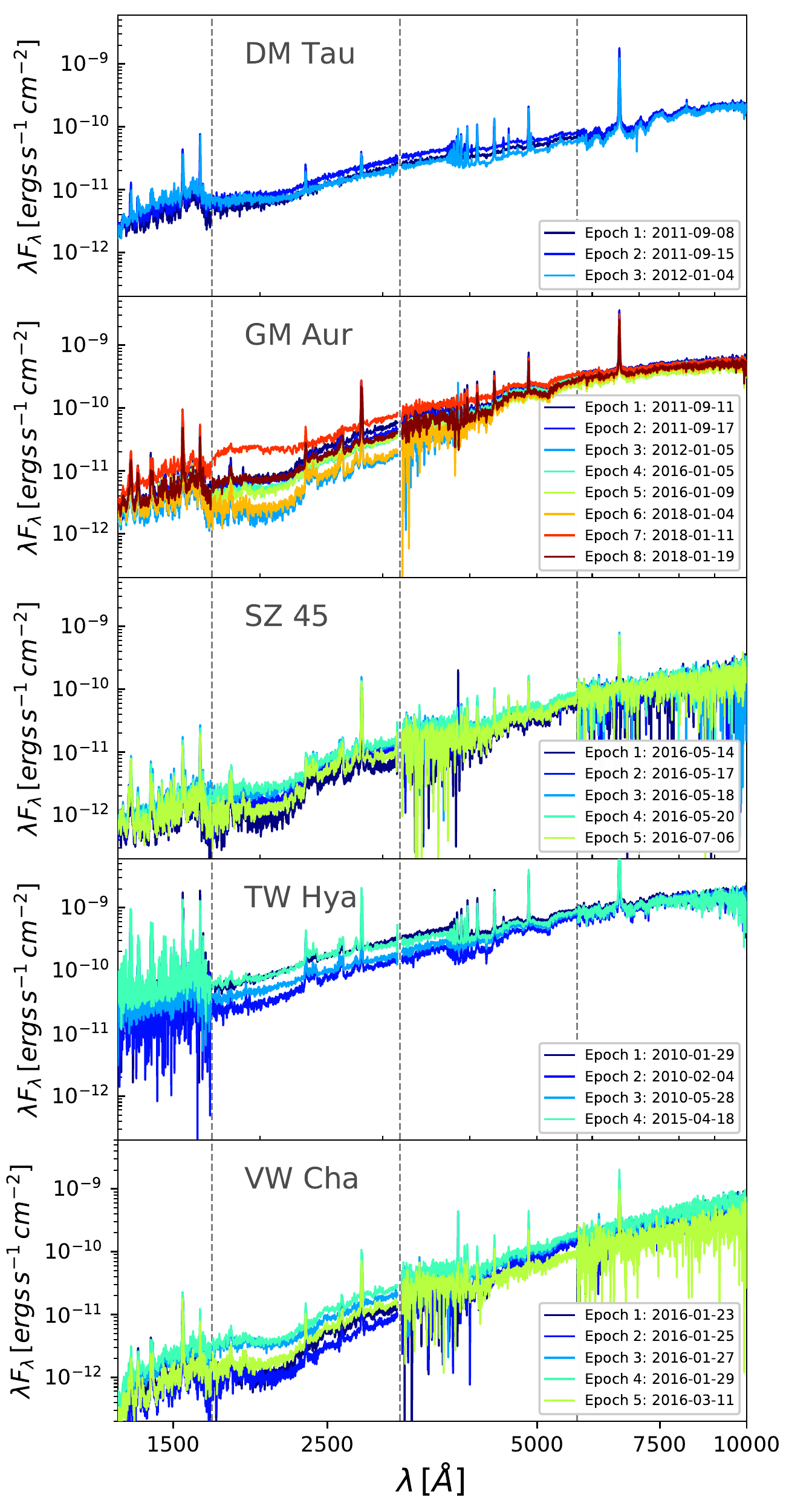}
    \caption{All 25 epochs of \textit{HST} STIS spectra with separate panels for each object. The grey dashed lines mark the edges of the STIS gratings.}
    \label{fig:allspectra}
\end{figure}

The observations presented here are taken from five separate \textit{HST} programs. A table summarizing the \textit{HST} proposals, dates, objects, and naming conventions used for epochs in this publication is shown in Table \ref{tab:hstobs}. The first three epochs of the GM Aur data (from Program 11608) have been previously studied by \citet{ingleby15} using similar techniques to this work. The WTTS used as templates in this analysis were observed as a part of \textit{HST} proposal 11616, PI: G. Herczeg. Fig. \ref{fig:allspectra} shows all of the spectra for each epoch stacked on top of each other.

\begin{deluxetable*}{ccccccc}
\tablecolumns{6}
\tablewidth{2.0\columnwidth} 
\tablecaption{Observation Log} 
\tablehead{Target & Epoch & Proposal & Date & Exposure & Notes \\ & & ID & & times [s] &}
\startdata
DM Tau & 1 & 2011-09-08 & 11608 & 25, 400, 821, 2994 &  \\
DM Tau & 2 & 2011-09-15 & 11608 & 25, 400, 821, 2994 &  \\
DM Tau & 3 & 2012-01-04 & 11608 & 25, 400, 821, 2994 &  \\
& & & & & &\\
GM Aur & 1 & 2011-09-11 & 11608 & 8, 35, 1231, 3020 &  \\
GM Aur & 2 & 2011-09-17 & 11608 & 8, 35, 1231, 3020 &  \\
GM Aur & 3 & 2012-01-05 & 11608 & 8, 35, 1231, 3020 &  \\
GM Aur & 4 & 2016-01-05 & 14048 & 8, 35, 1207, 2967 &  \\
GM Aur & 5 & 2016-01-09 & 14048 & 8, 35, 1207, 2967 &  \\
GM Aur & 6 & 2018-01-04 & 15165 & 8, 35, 1093, 2950 &  \\
GM Aur & 7 & 2018-01-11 & 15165 & 8, 35, 1093, 2950 &  \\
GM Aur & 8 & 2018-01-19 & 15165 & 8, 35, 1093, 2950 &  \\
& & & & & &\\
SZ 45 & 1 & 2016-05-14 & 14193 & 3, 45, 1549, 3298 &  \\
SZ 45 & 2 & 2016-05-17 & 14193 & 3, 45, 1549, 3298 &  \\
SZ 45 & 3 & 2016-05-18 & 14193 & 3, 45, 1549, 3298 &  \\
SZ 45 & 4 & 2016-05-20 & 14193 & 3, 45, 1549, 3298 &  \\
SZ 45 & 5 & 2016-07-06 & 14193 & 3, 45, 1549, 3298 &  \\
& & & & & &\\
TW Hya & 1 & 2010-01-29 & 11608 & 1, 20, 100, 150 & E140M \\
TW Hya & 2 & 2010-02-04 & 11608 & 1, 20, 100, 150 & E140M \\
TW Hya & 3 & 2010-05-28 & 11608 & 1, 20, 150, 548 & E140M \\
TW Hya & 4 & 2015-04-18 & 13775 & 1, 20, 150, 485 & E140M \\
& & & & & &\\
VW Cha & 1 & 2016-01-23 & 14193 & 2, 25, 1671, 3298 &  \\
VW Cha & 2 & 2016-01-25 & 14193 & 2, 25, 1671, 3298 &  \\
VW Cha & 3 & 2016-01-27 & 14193 & 2, 25, 1671, 3298 &  \\
VW Cha & 4 & 2016-01-29 & 14193 & 2, 25, 1671, 3298 &  \\
VW Cha & 5 & 2016-03-11 & 14193 & 2, 25, 1671, 3298 & Companion \\
& & & & & in slit &\\
HBC 427 & 1 & 2011-03-29 & 11616 & 60, 1475 & WTTS \\
RECX 1 & 1 & 2010-01-22 & 11616 & 14, 808 & WTTS \\
TWA 7 & 1 & 2011-05-05 & 11616 & 8, 1490 & WTTS \\
\enddata  
\label{tab:hstobs}
\tablecomments{FUV observations of TW Hya used the higher resolution E140M grating instead of the G140L grating that was used for the other objects to avoid saturation. A companion star landed on the slit during Epoch 5 of our VW Cha observations. The PIs for the proposals are as follows: 11608: N. Calvet, 13775: C. Espaillat, 14048: C. Espaillat, 14193: C. Espaillat, and 15165: C. Espaillat. Exposure times for the CTTS observations are listed in the following order: G750L, G430L, G230L, and G140L/E140M. The two listed exposure times for the WTTS are G230L and G140L observations, respectively.  More details on the observations are included in \S 2.3.}
\end{deluxetable*}

\subsubsection{De-fringing}
Fringe patterns were present in the longer wavelength regions of the NIR G750L spectra. These were removed by following the prescription presented in \citet{Goudfrooij98} using fringe flats and the STSDAS PyRAF tools. Further reduction of the observations followed the standard STScI \texttt{calstis} pipeline. 

\subsubsection{VW Cha}
One irregularity in the standard reduction process occurred while extracting the spectrum of Epoch 5 for VW Cha. Observations of VW Cha in the first four epochs were taken with similar slit position angles ($140-145^\circ$). The final epoch was taken with a $40^\circ$ deviation from this. The change in slit position caused a companion star to align with the slit. This companion is brighter in the NIR grating observation than VW Cha, which caused the companion to be extracted under the standard automated spectral extraction process. Using the STIS PyRAF tools, VW Cha was manually re-extracted from the 2D spectra with very little contamination from the companion. Although the companion is present in the 2D spectra of the G140L, G230L and the G430L observations, it is faint and VW Cha was correctly selected and extracted by the automated pipeline. All of the other observations do not show visible contributions from companions in the 2D spectra.

\subsubsection{TW Hya}
The FUV order for TW Hya was not taken with the G140L grating, and was instead done with the higher resolution E140M grating because it would saturate if observed with the lower resolution grating. In order to facilitate comparisons between other objects, the spectra were resampled onto the same wavelength grid as the G140L spectra using SpecTres, a Python spectral resampling tool \citep{carnall17}.

\section{MODEL}
The optical and NUV continua were modeled using an updated version of the \citet{calvet98} shock models. This model consists of three regions: the pre-shock region, the post-shock region, and the underlying photosphere which is heated by emission from the post-shock region. The shock is described by the strong shock fluid Rankine-Hugoniot jump conditions and the pre-shock density and velocity which can be used to solve for the structure of the post-shock region. Emission from the shock itself is not explicitly included due to the limitations imposed by the fluid approach taken by \citet{calvet98}. A fully kinetic approach would be required to reproduce the conditions and emission in the shock itself.

We have updated the treatment of the pre-shock and post-shock regions presented in \citet{calvet98}. The method used to model the heated photosphere remains unchanged. Additionally, we make the same assumption that flux from the undisturbed photosphere outside of the accretion columns is identical to that of non-accreting WTTS. Previous iterations of these models interpolated over a pre-solved grid of pre-shock and post-shock solutions. The updated model now solves for the conditions in these regions for each unique parameter combination.

\subsection{Post-shock Region}
\begin{figure}
    \centering
    \includegraphics[width = .9\linewidth]{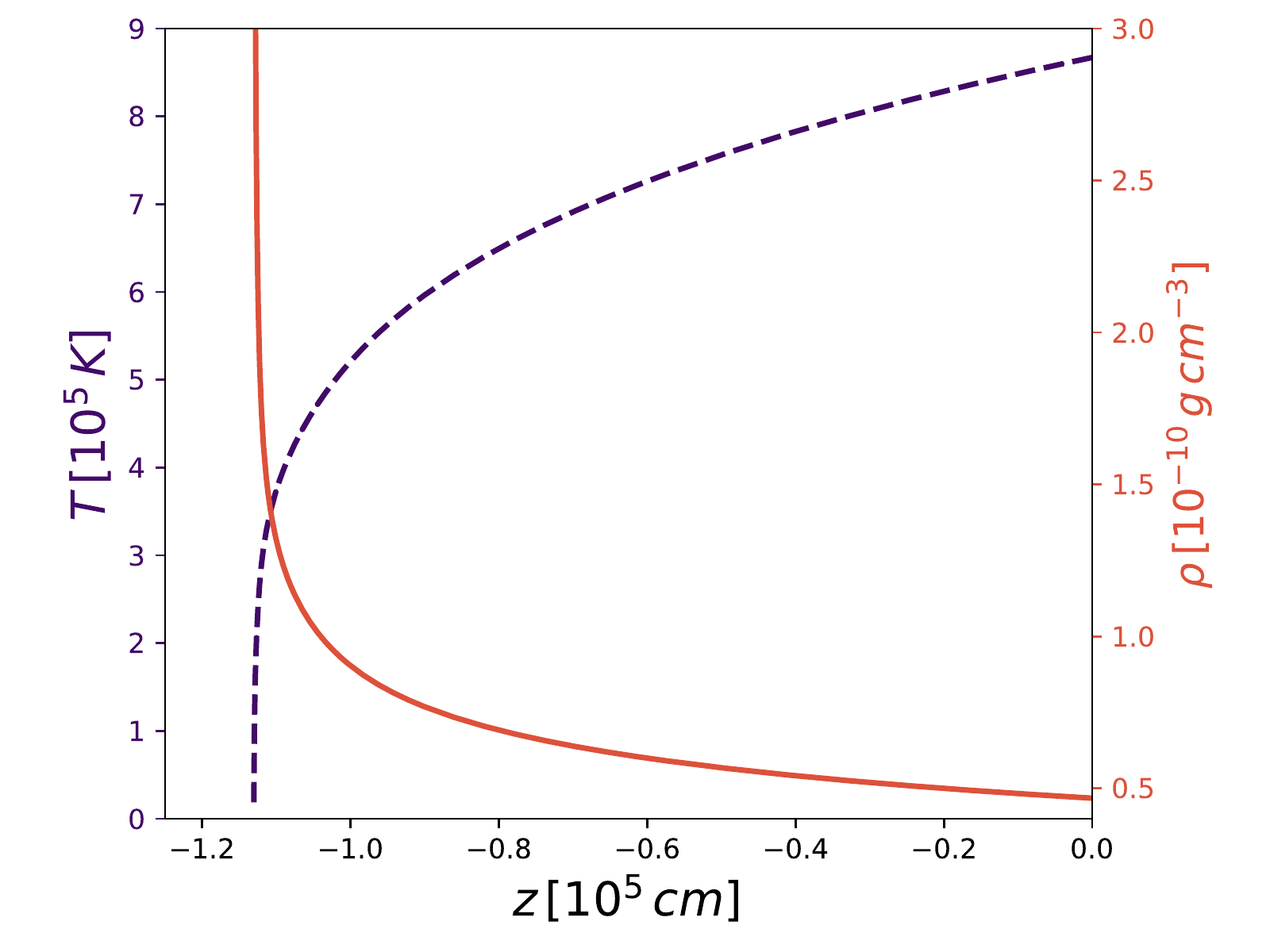}
    \includegraphics[width = .9\linewidth]{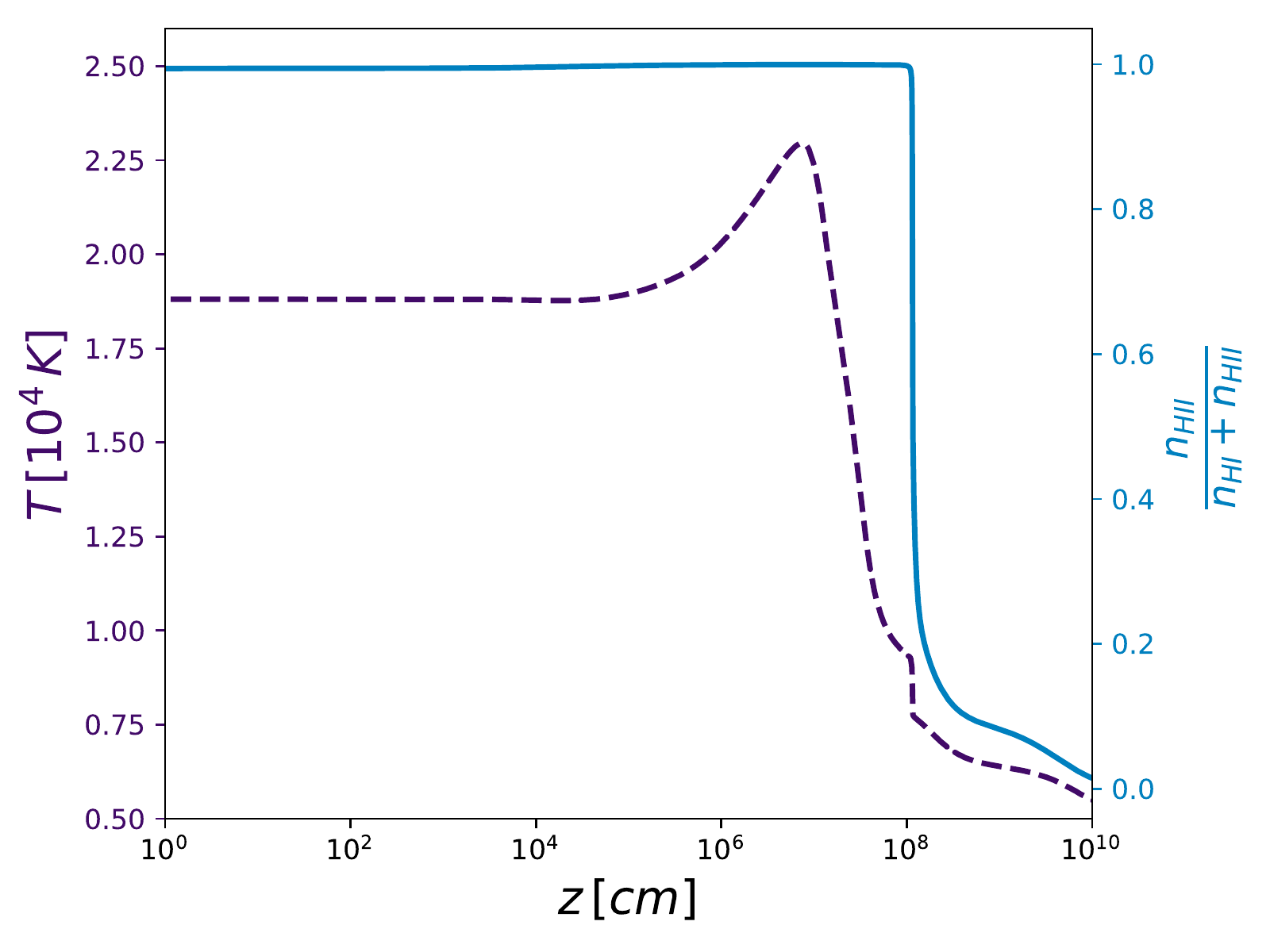}
    \caption{Temperature (dashed purple) and density (solid red) structure of the post-shock region (\textit{top}) and temperature and ionization fraction (solid blue) structure for the pre-shock regions (\textit{bottom}). These structures were solved using newly updated volume emissivities from Cloudy version c17.00 \citep{ferland17} for the case of GM Aur. This model was generated with an energy flux of $1\times10^{11} \, ergs \, s^{-1} \, cm^{-2}$. }
    \label{fig:shockstruct}
\end{figure}

The fluid equations governing the post-shock structure are solved using a 4th order Runge-Kutta solver. We have incorporated an adaptive mesh since the temperature in the post-shock region near the photosphere changes rapidly over small scales. Volume emissivities, emergent spectra, and optical depths for each cell were found by interpolating between a grid of models generated using Cloudy \citep[release c17.00;][]{ferland17} for a wide range of temperatures and densities. This grid covers $T = 5\times10^{3}$ to $1\times10^{7}$ K and $n = 10^{10}$ to $10^{17}\mbox{cm}^{-3}$ in logarithmic steps. The post-shock region is assumed to be optically thin, so the emission from each cell is summed to find the emergent radiation field. The emission is assumed to be isotropic, which results in the heated photosphere and the overlaying pre-shock region each receiving half of the total emission generated in the post-shock. The volume emissivities from the c17.00 release of Cloudy for plasma in this temperature regime are generally higher than the emissivities used in the original models \citep{raymond77}. The result of this is that the post-shock region is roughly a factor of 2 smaller than previous estimates produced using this model. Figure \ref{fig:shockstruct} shows an example the structure of the pre-shock and post-shock structures for GM Aur with these updates. 

\subsection{Pre-Shock Region}
The structure and emission, both towards and away from the stellar surface, of the pre-shock region is found using the built in radiative transfer methods of Cloudy \citep{ferland17}. The incident radiation field on the pre-shock region is the emission from the post-shock region. The density is fixed in this region to a value set by the energy flux of the column and the free-fall velocity. The outer boundary stopping condition is set when the depth of the modeled accretion column reaches $0.1 R_{\star}$ or when the temperature of the gas reaches 4000K. 

\subsection{Filling Factors}
These models solve for the outgoing emission from $1\,\mbox{cm}^{2}$ of the star while ignoring geometric viewing effects. Surface coverage for different energy flux columns is included using multiplicative filling factors, $f_i$. These filling factors represent the fraction of the star that is covered by accretion columns with that energy flux. By construction, an individual filling factor cannot be less than 0, and the sum of all filling factors cannot be greater than 1. Because these are simple multiplicative scaling factors, these filling factors can be changed after the radiative transfer modeling has finished. The emission is further scaled to account for the distance and radius of the star. 

\subsection{Photospheric emission}
We include emission from the undisturbed stellar photosphere using a WTTS with a similar spectral type as a template. Photospheric emission of CTTS is veiled by emission from accretion shocks \citep{hartigan89}. Veiling is a measure of how much shallower absorption lines look due to an additional continuum source.
The degree of spectral line veiling is often used to scale WTTS template spectra to the appropriate level for the CTTS.
Previous attempts at fitting continuum levels using the shock models of \citet{calvet98} often assume a constant value for veiling at V-band, $r_V$, between epochs \citep[e.g.,][]{ingleby13, ingleby15}, where $F_{V,WTTS} = F_{V,CTTS}/(1+r_V)$ since simultaneous high-resolution spectra can be difficult to obtain.
The goal of this work is to measure changes in the mass accretion rate, which should be highly correlated with veiling given that optical emission is produced by accretion shock models \citep{calvet98}. This means that if a constant V-band veiling is assumed, the inferred mass accretion rates measured will be systematically inaccurate. 

\begin{figure}
    \centering
    \includegraphics[width = .95\linewidth]{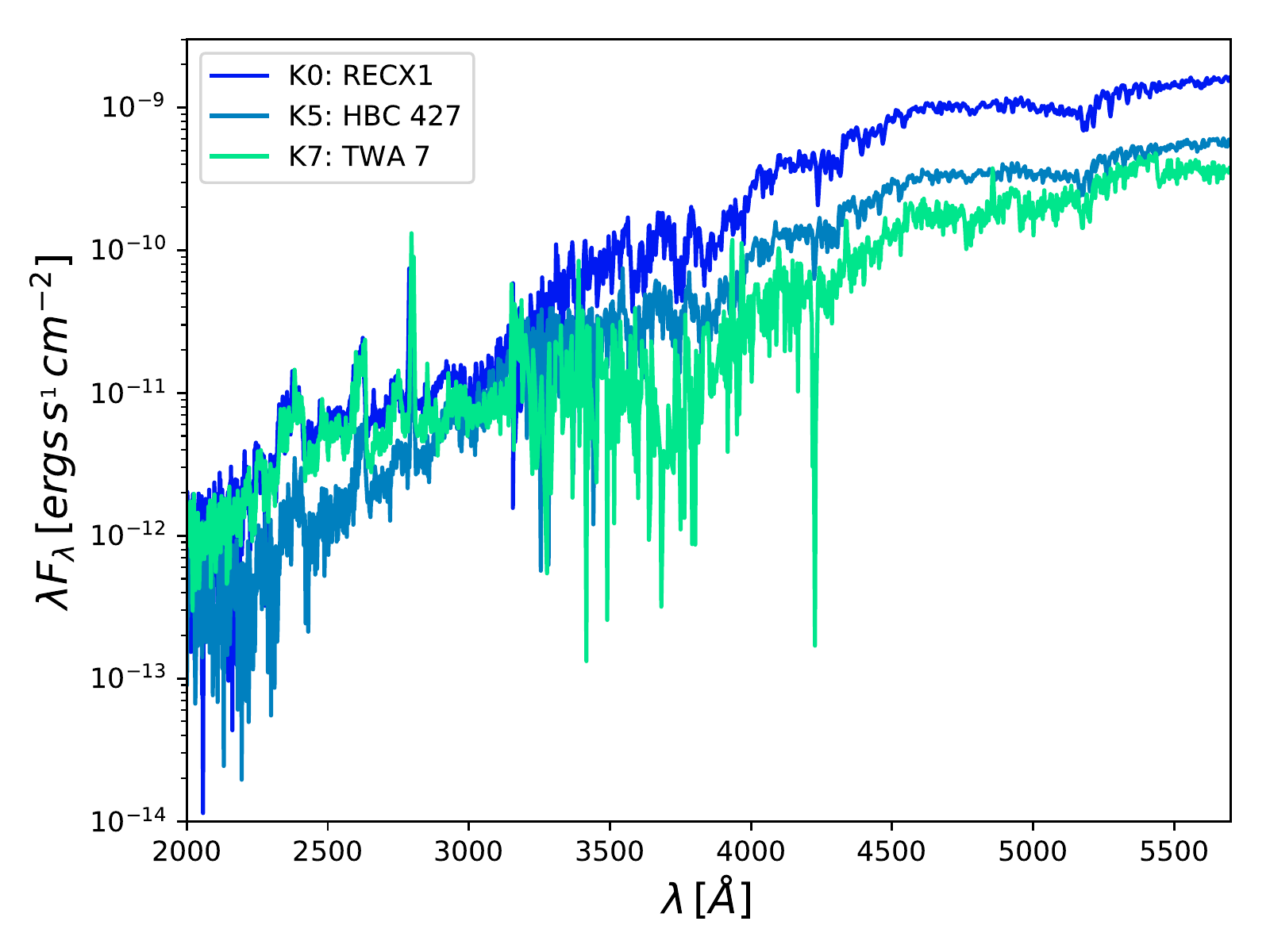}
    \caption{Weak-line T Tauri star spectra used as photospheric templates.}
    \label{fig:wtts}
\end{figure}

To approach this issue, we include the scaling of the undisturbed photospheric emission in our fitting as a parameter.
First, the WTTS template is reddened to the measured extinction of the CTTS. 
Next, the template is scaled by a factor, $s$, of the maximum flux at $5500\si{\angstrom}$ (roughly V-band) for all epochs of \textit{HST} STIS observations of that object. This approach fixes the emission from each $cm^2$ of the undisturbed photospheric flux of the star at a single level across all observational epochs.
By construction, $s$ should only take values between 0 and 1. 
The emission from this photospheric template is then multiplied by the fraction of the star that is not covered by accretion shocks ($1 - \sum_i f_i$). The WTTS photospheres used in this analysis can be seen in Figure \ref{fig:wtts}.

\begin{deluxetable*}{cccccccccccc}
\tablecolumns{13}
\tablewidth{2.0\columnwidth} 
\tablecaption{Stellar parameters} 
\tablehead{ Star & RA [J2000] & Dec. [J2000] & Distance [pc] &  SpT & T [K] & Mass [$M_{\odot}$] & Radius [$R_{\odot}$] & $A_v$ & WTTS & $i$ [deg.] & Ref.}
\startdata
DM Tau & 04:33:48.7 & +18:10:10.0 & $145.1\pm1.1$ & M2 & 3560 & 0.56 & 1.63 & 1.1 & TWA 7 & $34 ^{+2}_{-2}$ & [1,4] \\
GM Aur & 04:55:11.0 & +30:21:59.5 & $159.6\pm2.1$ & K5 & 4350 & 1.36 & 2.0 & 0.6 & RECX 1 & $52.77 ^{+0.05}_{-0.04}$ & [2,5] \\
TW Hya & 11:01:51.9 & -34:42:17.0 & $60.09\pm0.15$ & K7 & 4060 & 0.79 & 0.929 & 0.0 & HBC 427 & $7^{+3}_{-3}$ & [2,5] \\
SZ 45 & 11:17:37.0 & -77:04:38.1 & $188.4\pm0.9$ & M0.5 & 3780 & 0.85 & 1.78 & 0.7 & TWA 7 & -- & [2] \\
VW Cha & 11 08 01.5 & -77 42 28.9 & $190\pm5$ & K7 & 4060 & 1.24 & 3.08 & 1.9 & HBC 427 & -- &[3] \\
\enddata  
\label{table:cttsparams}
\tablecomments{Stellar parameters were collected from [1]: \citet{manara14}, and [2]: \citet{manara17} using the \citet{baraffe98} evolutionary tracks. Distances were taken from the GAIA DR2 \citep{gaia16, gaia18b} for all targets except VW Cha which did not have a parallax available, and instead uses the median distance for the Chamaeleon I star-forming region of 190 pc \citep{roccatagliata18}. Inclinations were collected when available from [3]: \citet{tripathi17}, [4]: \citet{macias18}, and [5]: \citet{andrews16}.}
\end{deluxetable*}
\begin{deluxetable}{ccccccc}
\tablewidth{2.0\columnwidth} 
\tablecaption{Weak-line T Tauri Stellar Parameters} 
\tablehead{ Star & RA [J2000] & Dec. [J2000] &  SpT & T [K] & $A_v$ & Ref.}
\startdata
    HBC 427 & 04:56:02 & +30:21:03 & K7 & 4060 & 0 & [2]\\
    RECX 1  & 08:36:56 & -78:56:45 & K5 & 4350 & 0 & [1]\\
    TWA 7   & 10:42:30 & -33:40:16 & M1 & 3720 & 0 & [3]
\enddata  
\label{tab:wttsparams}
\tablecomments{Sources for stellar parameters: [1]: \citet{luhman04}, [2]:\citet{kenyon95}, [3]:\citet{webb99}. Conversion from spectral type to temperature was done with the tables in \citet{kenyon95}.}
\end{deluxetable}

\subsection{Combining the model components}
Combining the components from the shock model and the undisturbed photosphere, the surface averaged outgoing flux corrected for extinction for a given wavelength for a single epoch can be written as 
\begin{equation}
F_{tot} = 10^{-0.4A_\lambda}\Big[\Big(1 - \sum^n_i f_i\Big) sF_{phot} + \sum^n_if_iF_i\Big]
\end{equation}
where each $i$ represents a different column density used in the construction of the model spectra and $n$ is the total number of contributing column densities (in our case, 3). $F_i$ is the sum of $F_{i,hp}$, the flux from the heated photosphere, and $F_{i,pre}$ the emission from the pre-shock region. $F_{phot}$ represents the emission arising from undisturbed photosphere, here scaled by $s$ and the fraction of the star not covered by accretion shocks. $A_\lambda$ is the extinction at each wavelength. Extinction is applied to the model using the extinction parameters for HD 29647 from \citet{whittet04} using the formulation from \citet{fitzpatrick88} and \cite{fitzpatrick90}.
Emission from the post-shock region is not explicitly included in this equation because the outgoing emission from the post-shock forms the incident radiation field for the pre-shock region and the heated photosphere. 

\section{ANALYSIS AND RESULTS}

\subsection{Stellar Parameters}
Stellar parameters for the CTTS for this sample are included in Table \ref{table:cttsparams}. Relevant stellar parameters for WTTS used as templates are listed in Table \ref{tab:wttsparams}. These parameters were collected from the literature \citep{manara14, manara17}, and were calculated by those authors using the \citet{baraffe98} evolutionary models. Distances were calculated using parallax measurements from the Gaia Data Release 2 \citep{gaia16,gaia18b}. Stellar radii from these sources were scaled to be self consistent with the new GAIA distances. Stellar temperature estimates were found using the relationships between $T_{eff}$ and spectral type for pre-main sequence stars in \citet{kenyon95}.

\subsection{Fitting the NUV/optical continuum}
\begin{figure*}
    \centering
    \includegraphics[width = .9\linewidth]{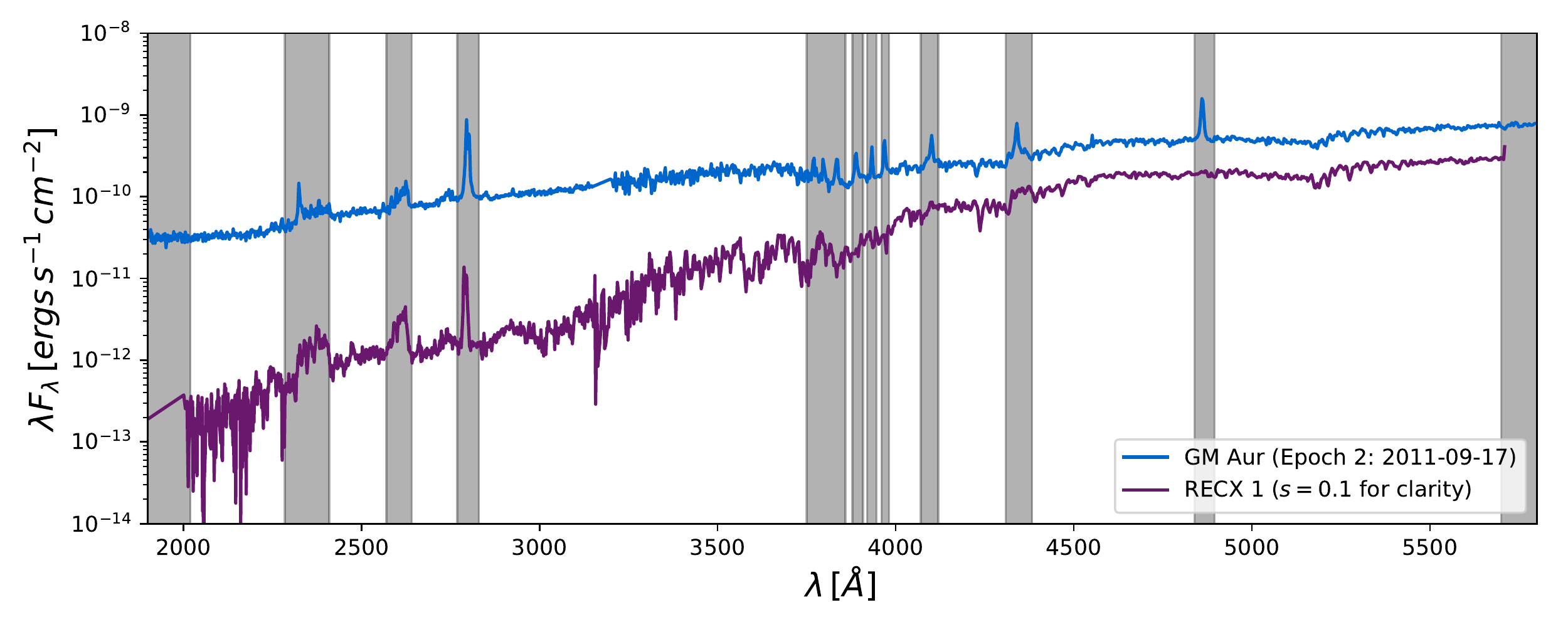}
    \caption{An accreting star (GM Aur; blue) with an example non-accreting template photosphere (RECX 1; purple).
    Grey shaded areas show regions that were not included when fitting filling factors, and were selected to avoid spectral features that were not included in the modeling process. The WTTS is shown scaled to an $s$ value of 0.1 for clarity, and the CTTS spectra has been de-reddened.}
    \label{fig:mask}
\end{figure*}

We model the excess NUV/optical continuum emission with the sum of emission from accretion columns with three different energy fluxes, $F_i$, in logarithmic steps: $1\times10^{10}, 1\times10^{11}, 1\times10^{12} \, ergs \, s^{-1}\, cm^{-2}$ along with the scaled undisturbed photosphere template. The velocity at the shock is assumed to be the free-fall velocity, and the energy of the flow is assumed to be solely kinetic. This causes the energy flux to be directly related to the density of the column. Regions with strong line emission were masked (the mask can be seen in Figure \ref{fig:mask}) and were not included when fitting models to data.  To fit our data, we employ a Bayesian Monte Carlo Markov Chain (MCMC) approach, which is described in the following sections.

\subsubsection{Likelihood}
Under the assumption that the emission from the undisturbed photosphere does not change between observations, we can treat $s$ as a single parameter that affects all of the epochs simultaneously for a single object.
Fractional uncertainty in the model, $w$, is included as a nuisance parameter. This uncertainty is added in quadrature with the measurement uncertainty when calculating the likelihood of a given model. 
Assuming Gaussian and independent uncertainties, the likelihood function is written
\begin{equation}
\begin{split}
    P(D|\{f_{ij}\}, s, w, I) = \\
    \prod^q_j\prod^{r_j}_k \frac{1}{\sqrt{2\pi(\sigma_{jk}^2 + w^2F_{jk}^2})} 
    \cdot \exp\Big[-\frac{1}{2} \frac{(F_{jk} - D_{jk})^2}{(\sigma_{jk}^2 + w^2F_{jk}^2)}\Big]
\end{split}
\end{equation}
where $j$ indexes over $q$ observational epochs, $k$ indexes over $r_j$ data points for each epoch, $i$ again indexes over the different accretion column energy fluxes, $\sigma_{jk}$ is the measurement uncertainty in each point from the \textit{HST} pipeline, $\{f_{ij}\}$ represents the complete set of filling factors for all epochs, $D$ is the data, and $F$ is the model (which is a function of $\{f_{ij}\}$ and $s$, see Eqn. 1).
The number of free parameters in our fit scales as $2 + 3q$ since we have chosen to include contributions from three column densities and $s$ and $w$ remain single parameters for all epochs. 

\subsubsection{Priors}
Priors for each filling factor, $s$, and $w$ were included in this analysis. Priors allow us to include prior known information about model parameters when calculating posterior distributions.
Each parameter was evaluated in log-space, and had a cut-off placed at 0 which limits the values of the parameters in regular-space to be between 0 and 1. For the filling factors, this prior removes the nonphysical possibility of both negative emission and surface coverage larger than the surface of the star. In the case of $s$, the lower bound enforces that the photospheric contribution is not negative, while the upper bound is set because we would not expect the undisturbed photosphere to be brighter than a model that includes additional energy input from the accretion shock. In addition to these cut-offs, we include a prior consisting of a Gaussian centered at the expected value for $s$ based on previous shock modeling by \citet{manara14, manara16a} with a width of 0.1 in $s$ space. Limiting $w$ between 0 and 1 ensures that the uncertainity is positive, and that the emission from the shock model is also positive within the uncertainties. In practice, $w$ does not approach 1.

\subsubsection{Sampling the posterior}
\begin{figure*}
    \centering
    \includegraphics[width = .98\linewidth]{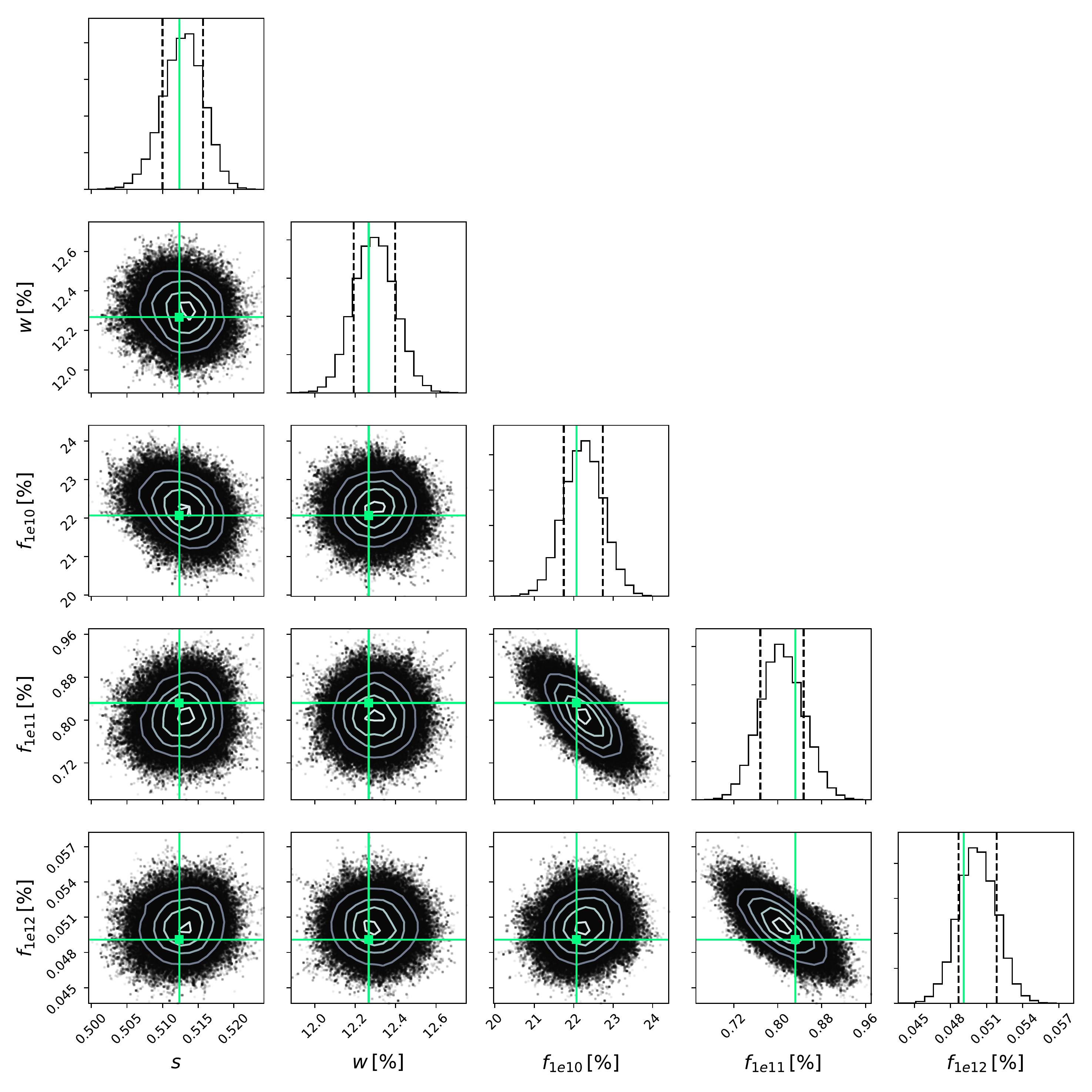}
    \caption{2D marginalized posteriors of the photospheric scaling factor $s$, fractional model uncertainty $w$, and filling factors for Epoch 2 of the GM Aur observations (written as percentages). The most likely model is marked in green. Note that filling fractions for all epochs for each object were fit simultaneously and that only a subset parameters included in our analysis is shown here as an example.}
    \label{fig:corner}
\end{figure*}

To sample the posterior, we employ the widely-used ensemble sampler \texttt{emcee} \citep{foreman-mackey13} which uses the \citet{goodman10} affine-invarient sampling algorithm. 
We used 100 walkers to sample the parameter space for each object. A generous burn-in period before convergence was removed from each walker for each object. 
Fig. \ref{fig:corner} shows 2D marginalized posteriors for the subset of the parameters representing Epoch 2 of the GM Aur observations. Marginalized posteriors represent the probability of drawing a particular value of a parameter without reference to parameters that have been marginalized over. In Fig. \ref{fig:corner}, all of the filling factors for the other epochs have been marginalized over. 
Anti-correlations between $f_{1e10}$ and $f_{1e11}$ and between $f_{1e11}$ and $f_{1e12}$ appear due to some degree of degeneracy within the models. 
We find that typical values of the nuisance parameter, $w$, which represents the fractional uncertainty in the model generally falls in the range of 10\% to 20\% for each object.

\subsubsection{Mass accretion rates}
Mass accretion rates for each epoch can be calculated by summing up the contributions from all the columns via
\begin{equation}
    \dot{M} = 8\pi\Big(\frac{R_\star}{v_{ff}}\Big)^2 \sum_i^n f_iF_i
\end{equation}
where $V_{ff}$ is the free-fall velocity from the inner edge of the gas disk, $R_{in}$, written as 
\begin{equation}
    V_{ff} = \sqrt{\frac{2GM_\star}{R_\star}} \sqrt{1-\frac{R_\star}{R_{in}}}.
\end{equation}
Here, $R_{in}$ is assumed to be 5 stellar radii. 
It is important to note that the mass accretion rates reported here are calculated under the assumption that the surface coverage of the columns is identical on the far side of the star. In addition, these models are 1D and ignore geometrical effects which would require a much more detailed analysis. 

Mass accretion rates and filling factors are reported in Table \ref{tab:modelfits}. The median of the posterior distribution is reported for each of these parameters. Reported uncertainties are the difference between the median and the $16^{th}$ and $84^{th}$ percentiles (roughly equivalent to $1\sigma$ uncertainties for Gaussian distributions). Values of veiling for each epoch at V-band, parameterized by $r_V$, are also presented in Table \ref{tab:modelfits}. 
 The degree of veiling is related to the parameter $s$, the photospheric scaling factor described earlier, and the amount of excess continuum flux from the accretion shocks. 
One of the V-band veiling values for VW Cha was found to be negative, which is unusual. This peculiar epoch is further discussed in \S 5.2.5.

\begin{deluxetable*}{r|cccccccccc}
\centering
\tablecolumns{11}
\tablecaption{Mass Accretion Rates and Filling Factors}
\tablewidth{2.0\columnwidth} 
\tablehead{Object & Epoch & $\dot{M}$  & $f_{1E10}$  & $f_{1E11}$  & $f_{1E12}$ & $r_V$}
\startdata
DM Tau & 1 & $2.770^{+0.026}_{-0.025}$ & $0.211^{+0.007}_{-0.007}$ & $0.000066^{+0.00004}_{-0.000015}$ & $0.003638^{+0.000027}_{-0.000026}$ & $1.41^{+0.05}_{-0.05}$ \\
DM Tau & 2 & $3.582^{+0.029}_{-0.029}$ & $0.259^{+0.008}_{-0.008}$ & $0.000069^{+0.00005}_{-0.000018}$ & $0.00485^{+0.00003}_{-0.00003}$ & $1.81^{+0.06}_{-0.05}$ \\
DM Tau & 3 & $2.011^{+0.028}_{-0.028}$ & $0.036^{+0.008}_{-0.008}$ & $0.000059^{+0.000024}_{-0.000010}$ & $0.003808^{+0.000027}_{-0.000027}$ & $1.06^{+0.04}_{-0.04}$ \\
GM Aur & 1 & $1.546^{+0.011}_{-0.011}$ & $0.214^{+0.006}_{-0.006}$ & $0.0164^{+0.0005}_{-0.0005}$ & $0.000459^{+0.000019}_{-0.000018}$ & $0.893^{+0.011}_{-0.011}$ \\
GM Aur & 2 & $1.291^{+0.010}_{-0.010}$ & $0.223^{+0.005}_{-0.005}$ & $0.0081^{+0.0004}_{-0.0004}$ & $0.000502^{+0.000016}_{-0.000016}$ & $0.847^{+0.011}_{-0.010}$ \\
GM Aur & 3 & $0.660^{+0.007}_{-0.007}$ & $0.142^{+0.003}_{-0.003}$ & $0.00342^{+0.00015}_{-0.00015}$ & $0.0000462^{+0.0000013}_{-0.0000006}$ & $0.643^{+0.010}_{-0.009}$ \\
GM Aur & 4 & $1.021^{+0.009}_{-0.009}$ & $0.183^{+0.004}_{-0.004}$ & $0.0057^{+0.0003}_{-0.0003}$ & $0.000390^{+0.000013}_{-0.000013}$ & $0.690^{+0.010}_{-0.010}$ \\
GM Aur & 5 & $0.768^{+0.008}_{-0.008}$ & $0.044^{+0.004}_{-0.004}$ & $0.0127^{+0.0003}_{-0.0003}$ & $0.000395^{+0.000013}_{-0.000013}$ & $0.303^{+0.008}_{-0.007}$ \\
GM Aur & 6 & $0.564^{+0.007}_{-0.007}$ & $0.087^{+0.003}_{-0.003}$ & $0.00528^{+0.00021}_{-0.00020}$ & $0.000147^{+0.000008}_{-0.000008}$ & $0.428^{+0.008}_{-0.008}$ \\
GM Aur & 7 & $1.961^{+0.012}_{-0.012}$ & $0.303^{+0.005}_{-0.005}$ & $0.00009^{+0.00008}_{-0.00003}$ & $0.002326^{+0.000022}_{-0.000022}$ & $0.950^{+0.011}_{-0.011}$ \\
GM Aur & 8 & $0.979^{+0.009}_{-0.009}$ & $0.125^{+0.005}_{-0.005}$ & $0.0070^{+0.0004}_{-0.0004}$ & $0.000726^{+0.000017}_{-0.000016}$ & $0.551^{+0.009}_{-0.009}$ \\
SZ 45 & 1 & $0.923^{+0.015}_{-0.015}$ & $0.172^{+0.006}_{-0.006}$ & $0.00452^{+0.00028}_{-0.0003}$ & $0.000057^{+0.000009}_{-0.000007}$ & $1.71^{+0.10}_{-0.10}$ \\
SZ 45 & 2 & $1.187^{+0.018}_{-0.018}$ & $0.191^{+0.007}_{-0.007}$ & $0.0064^{+0.0005}_{-0.0005}$ & $0.000321^{+0.000018}_{-0.000017}$ & $2.08^{+0.12}_{-0.11}$ \\
SZ 45 & 3 & $1.490^{+0.019}_{-0.019}$ & $0.247^{+0.008}_{-0.008}$ & $0.0060^{+0.0006}_{-0.0006}$ & $0.000528^{+0.000024}_{-0.000023}$ & $2.70^{+0.14}_{-0.13}$ \\
SZ 45 & 4 & $1.639^{+0.020}_{-0.020}$ & $0.300^{+0.008}_{-0.008}$ & $0.0034^{+0.0006}_{-0.0006}$ & $0.000631^{+0.000026}_{-0.000025}$ & $2.86^{+0.15}_{-0.14}$ \\
SZ 45 & 5 & $1.157^{+0.018}_{-0.017}$ & $0.204^{+0.007}_{-0.007}$ & $0.0058^{+0.0004}_{-0.0004}$ & $0.000172^{+0.000015}_{-0.000014}$ & $2.27^{+0.13}_{-0.11}$ \\
TW Hya & 1 & $0.330^{+0.004}_{-0.004}$ & $0.275^{+0.014}_{-0.016}$ & $0.0113^{+0.0011}_{-0.0011}$ & $0.001319^{+0.00003}_{-0.000028}$ & $0.962^{+0.020}_{-0.022}$ \\
TW Hya & 2 & $0.1384^{+0.0028}_{-0.003}$ & $0.083^{+0.009}_{-0.010}$ & $0.0093^{+0.0006}_{-0.0006}$ & $0.000427^{+0.000013}_{-0.000014}$ & $0.500^{+0.015}_{-0.017}$ \\
TW Hya & 3 & $0.2206^{+0.0028}_{-0.004}$ & $0.256^{+0.008}_{-0.013}$ & $0.0007^{+0.0009}_{-0.0006}$ & $0.000852^{+0.000016}_{-0.000023}$ & $0.695^{+0.017}_{-0.019}$ \\
TW Hya & 4 & $0.262^{+0.005}_{-0.004}$ & $0.153^{+0.016}_{-0.013}$ & $0.0120^{+0.0010}_{-0.0012}$ & $0.00140^{+0.00004}_{-0.00003}$ & $0.818^{+0.019}_{-0.020}$ \\
VW Cha & 1 & $8.5^{+0.3}_{-2.0}$ & $0.25^{+0.03}_{-0.25}$ & $0.0016^{+0.012}_{-0.0014}$ & $0.00303^{+0.00008}_{-0.00008}$ & $0.441^{+0.022}_{-0.021}$ \\
VW Cha & 2 & $7.46^{+0.20}_{-0.20}$ & $0.324^{+0.017}_{-0.018}$ & $0.00010^{+0.00011}_{-0.00004}$ & $0.00180^{+0.00005}_{-0.00004}$ & $0.408^{+0.021}_{-0.020}$ \\
VW Cha & 3 & $15.1^{+0.3}_{-0.3}$ & $0.386^{+0.029}_{-0.028}$ & $0.00014^{+0.00019}_{-0.00008}$ & $0.00638^{+0.00009}_{-0.00010}$ & $0.747^{+0.027}_{-0.025}$ \\
VW Cha & 4 & $19.9^{+0.4}_{-0.4}$ & $0.54^{+0.04}_{-0.04}$ & $0.00027^{+0.0006}_{-0.00019}$ & $0.00804^{+0.00013}_{-0.00012}$ & $0.941^{+0.03}_{-0.028}$ \\
VW Cha & 5 & $6.91^{+0.06}_{-0.05}$ & $0.0006^{+0.004}_{-0.0005}$ & $0.00010^{+0.00011}_{-0.00004}$ & $0.00465^{+0.00004}_{-0.00004}$ & $-0.074^{+0.014}_{-0.013}$ \\
\enddata
\tablecomments{Mass accretion rates are presented in units of $10^{-8} M_\odot\,yr^{-1}$ Filling factors are unitless, representing the fraction of the star covered by shocks with that energy flux. Reported values are the median of the distributions while the uncertainties are the difference between the median and the $16^{th}$ and $84^{th}$ percentile (similar to $1\sigma$ uncertainties). The negative values of veilings for VW Cha are discussed in depth in \S 5.2.5. 
It is important to note that uncertainties reported here only reflect the width of the 1D marginalized posterior as found from our MCMC analysis, and do not take other systematic effects into account (e.g., extinction, optical depth effects, evolutionary track). We find that typically a change in $A_v$ of 0.5 leads to systematic shifts in the absolute value of $\dot{M}$ by approximately a factor of 2, while relative shifts between epochs remain around $10\%$.}
\label{tab:modelfits}
\end{deluxetable*}

\subsection{NUV and FUV line luminosities}

Correlations have previously been found between NUV and FUV emission features and the mass accretion rate \citep[e.g.,][]{calvet04, ingleby13}.  Repeated observations of the same stars helps us separate trends that might appear due to systematic effects from those caused by changes in mass accretion rate.  Here we measure line luminosities for several NUV and FUV emission lines and $H_\alpha$ (Table \ref{tab:linelum}) by subtracting a continuum and directly integrating the feature across the relevant wavelength range. Kinematic information about the pre-shock region is also carried by these lines \citep{ardila13}, but the spectra presented here lack the resolution to do a detailed line shape study.

\begin{figure}
    \centering
    \includegraphics[width = .9\linewidth]{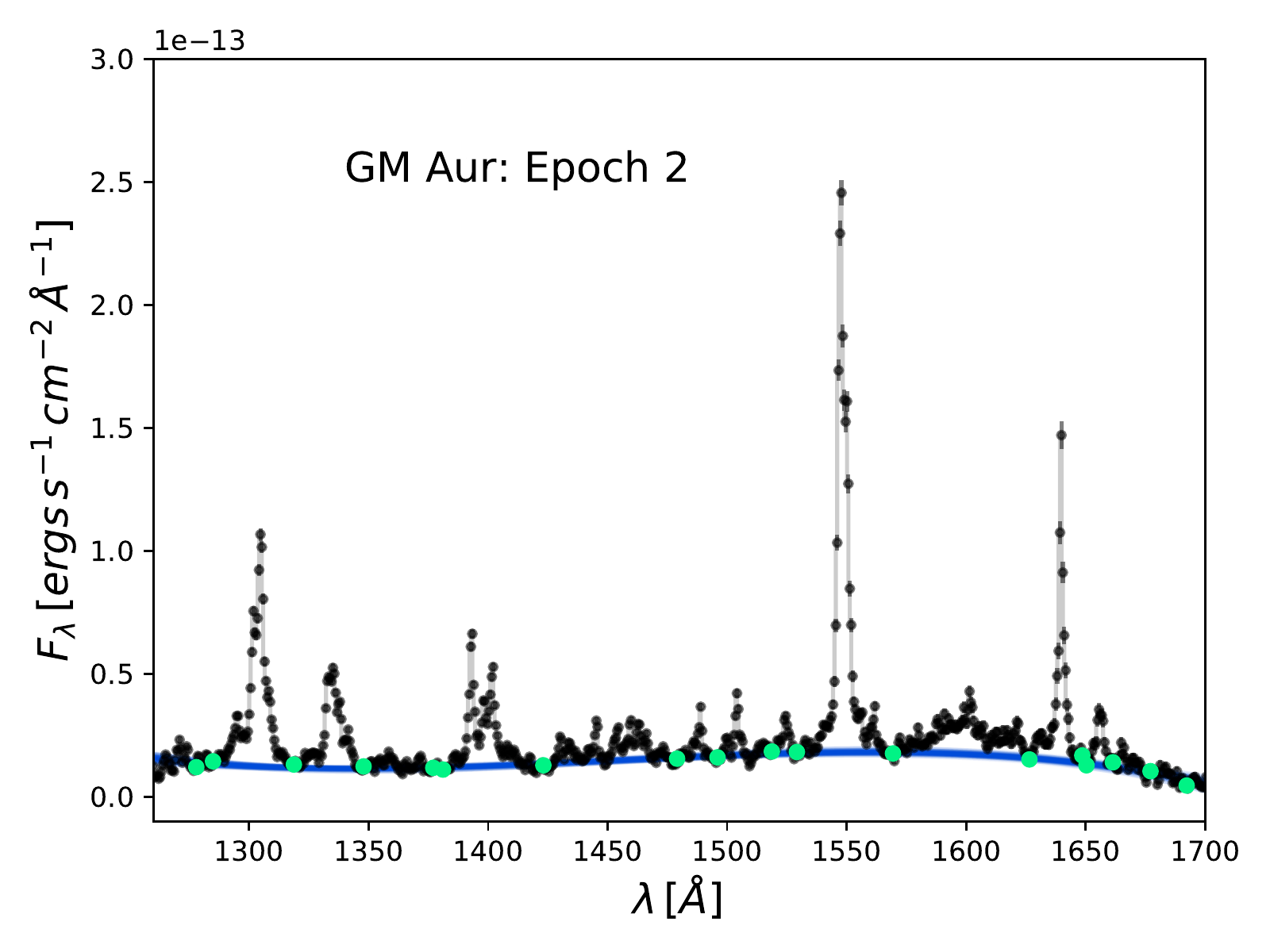}
    \caption{FUV (MAMA G140L) spectra for GM Aur Epoch 2 observations in black shown as an example of the continuum fitting process. Points selected as being representative of the continuum level are in green. A representative sample (100 models) drawn from the posterior for a 3rd degree polynomial fit are shown in blue. The width of the posterior is representative of the uncertainty in our fit of the continuum.}
    \label{fig:FUV_continuum}
\end{figure}

The continuum was fit with a 3rd degree polynomial. Although our model produces physically motivated continuum estimates, polynomial fits allow for more accurate representations of the observed continuum level near emission lines. In addition, the FUV was not included when fitting the filling fractions due to the number of strong spectral features, and thus should not be used for continuum subtraction with the shock model. Separate polynomials were used to fit the FUV and the NUV wavelength ranges for each observation to help ensure an accurate continuum subtraction. The continuum level was estimated by eye by selecting a set of points for each observation that appeared to be representative of the continuum level across the spectra. About 15 points were selected for both the FUV and NUV spectra. These points were then fit with basic linear regression techniques.
The resulting posterior distribution describes the probability of a model with a given set of parameters based on the available data and the prior. By randomly sampling the posterior and computing models, we can construct a set of models that begins to resemble the posterior placed in wavelength space. The width of these random draws from the posterior distribution is representative of the uncertainty in the continuum fit. Given the plethora of spectral lines, this approach appears to give a reasonable estimate of the uncertainty in the true level of the continuum. To ensure that the effect of line contamination in our continuum estimates is minimal, we compared our continuum fit to the non-convolved high resolution observations of TW Hya and found good agreement between the two. Additional tests consisting of resampling the data at resolutions between the COS resolution and STIS resolution with higher signal-to-noise ratios returned similar results.
An example of the selected points and a representative set of draws from the posterior for the FUV wavelengths for Epoch 2 of the GM Aur observations is shown in Fig. \ref{fig:FUV_continuum}.

Line luminosities and uncertainties were measured through a Monte Carlo approach. First, each data point in the relevant wavelength range was randomly re-sampled from a Gaussian centered on the measured value with width given by the measurement uncertainties from the \textit{HST} STIS pipeline. Next, a draw from the posterior for the continuum is selected and subtracted, and the line is integrated using a trapezoidal integration scheme. This was repeated many times, resulting in a posterior distribution for the line luminosity.  Due to line blending by the low resolution nature of the spectra, multiplet lines were measured and reported as individual spectral features by integrating over a narrow wavelength range covering the width of the feature. This includes the bright doublet lines $Mg_{II}$ and $C_{IV}$.

\subsection{Correlations with Observables}
\begin{figure*}
    \centering
    \includegraphics[width = .97\linewidth]{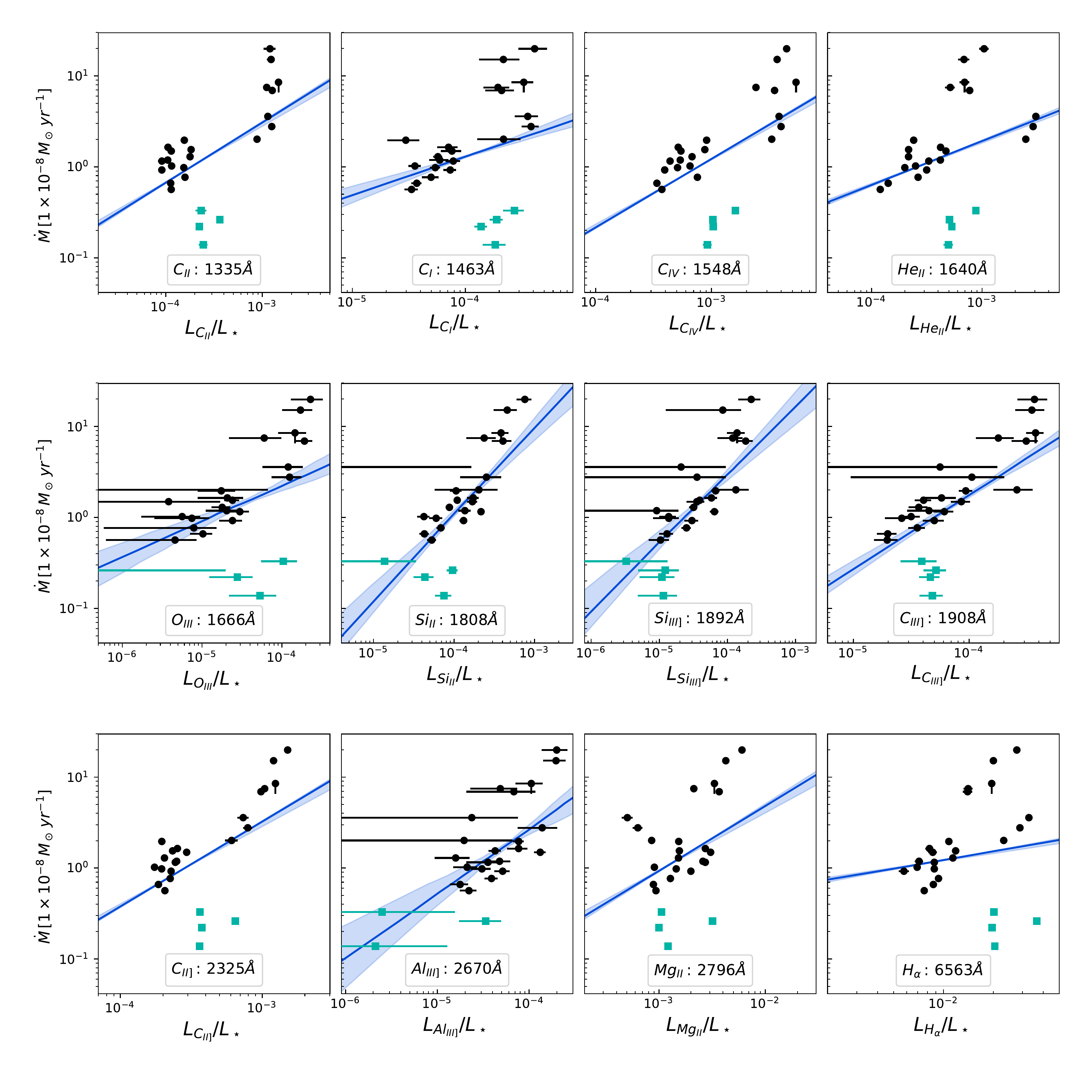}
    \caption{Log-linear fits to emission line luminosities and $\dot{M}$. The line luminosities have been divided by the stellar luminosity. The black dots represent observations for all of the objects except for TW Hya, which is shown as teal squares. TW Hya may be an outlier for several lines due to its viewing geometry. The log-linear fits shown here, whose coefficients are listed in Table \ref{tab:pearson}, include the points from TW Hya during the fitting process. The error bars in this figure reflect the values presented in Table \ref{tab:modelfits} and Table \ref{tab:linelum}.}
    \label{fig:big_corr}
\end{figure*}

Plotting luminosities for many of the readily distinguishable lines in our spectra against mass accretion rate reveals relatively strong correlations. These correlations were quantified using the Pearson Correlation Coefficient ($\rho$) in Table \ref{tab:pearson}. As a reminder, a perfect positive correlation will have a $\rho$ of 1. 
Fig. \ref{fig:big_corr} shows the line luminosities normalized by the stellar luminosity against $\dot{M}$.

We found that many of the NUV lines had the strongest correlations with the mass accretion rate. In particular, $Si_{II}, \, Si_{III]} \, C_{III]}, \, Al_{III]}$ all had values of $\rho$ greater than 0.7, with $Si_{III]}$ having the strongest correlation. 
$C_{IV}$ was found to have the strongest correlation between any of the FUV lines and the mass accretion rate, with a median $\rho$ of 0.620. $C_{IV}$ is quite bright, and has been previously observed and characterized in higher resolution studies \citep{ardila13}. This makes it useful as a mass accretion rate indicator when simultaneous NUV - optical measurements are not available \citep[e.g., spectra from the Cosmic Origins Spectrograph][]{green12}. Other possible options for relating FUV lines to $\dot{M}$ include $C_{II}$ and $O_{III}$. One possible explanation for the changes in line luminosity is contamination from the variable continuum level in our line measurements, especially in the narrow partially forbidden lines. 
However, the regions chosen for continuum subtraction were selected for each epoch individually with care to avoid line centers and wings while selecting points across the entire wavelength range (see Fig. \ref{fig:FUV_continuum} for an example). This makes it unlikely that our continuum extraction would be biased towards over predicting line fluxes for higher accretion rates.

In addition to correlations, we produced log-linear relationships between accretion rate and line strength with slope $m$ and offset $b$ which are also listed in Table \ref{tab:pearson}. These relationships take the form 
\begin{equation}
    \log_{10}\big(\frac{\dot{M}}{10^{-8}M_\odot \, yr^{-1}}\big) = m\cdot \log_{10}\big( \frac{L_{line}}{L_\star}\big) + b.
\end{equation}
These fits are shown plotted on top of the data in Fig. \ref{fig:big_corr}.  Uncertainties on both the fit and the correlation coefficient were found using a Monte Carlo approach using draws from the model fit posterior and the line luminosity posterior. The scatter around the lines in some cases is large relative to uncertainties in the fit, suggesting systematic effects outside measurement uncertainties. This is somewhat expected given the simplicity of this approach.
We find that line luminosities for individual objects follow relationships that are generally closer to monotonically increasing functions of mass accretion rate as expected. 
\begin{deluxetable}{ccccc}
\tablecolumns{10}
\tablewidth{2.0\columnwidth} 
\tablecaption{Pearson correlation coefficients and log-linear fits for FUV/NUV lines} 
\tablehead{Line & $\lambda \, [\si{\angstrom}]$ & $\rho$ & $m$ & $b$}
\startdata
$C_{II}$ & $1335\AA$ & $0.619^{+0.013}_{-0.018}$ & $0.660^{+0.018}_{-0.05}$ & $2.46^{+0.06}_{-0.17}$ \\
$C_{I}$ & $1463\AA$ & $0.32^{+0.05}_{-0.07}$ & $0.39^{+0.08}_{-0.07}$ & $1.7^{+0.3}_{-0.3}$ \\
$C_{IV}$ & $1548\AA$ & $0.620^{+0.012}_{-0.017}$ & $0.758^{+0.014}_{-0.05}$ & $2.36^{+0.04}_{-0.17}$ \\
$He_{II}$ & $1640\AA$ & $0.376^{+0.023}_{-0.021}$ & $0.480^{+0.029}_{-0.03}$ & $1.72^{+0.09}_{-0.11}$ \\
$O_{III}$ & $1666\AA$ & $0.48^{+0.11}_{-0.07}$ & $0.41^{+0.09}_{-0.11}$ & $2.0^{+0.4}_{-0.5}$ \\
$Si_{II}$ & $1808\AA$ & $0.77^{+0.05}_{-0.09}$ & $0.93^{+0.10}_{-0.17}$ & $3.8^{+0.4}_{-0.6}$ \\
$Si_{III]}$ & $1892\AA$ & $0.78^{+0.08}_{-0.14}$ & $0.74^{+0.13}_{-0.18}$ & $3.4^{+0.6}_{-0.8}$ \\
$C_{III]}$ & $1908\AA$ & $0.71^{+0.05}_{-0.07}$ & $0.80^{+0.07}_{-0.10}$ & $3.47^{+0.26}_{-0.4}$ \\
$C_{II]}$ & $2325\AA$ & $0.595^{+0.012}_{-0.020}$ & $0.933^{+0.022}_{-0.09}$ & $3.31^{+0.08}_{-0.3}$ \\
$Al_{III]}$ & $2670\AA$ & $0.69^{+0.09}_{-0.10}$ & $0.74^{+0.13}_{-0.19}$ & $3.4^{+0.6}_{-0.8}$ \\
$Mg_{II}$ & $2796\AA$ & $0.408^{+0.013}_{-0.025}$ & $0.71^{+0.04}_{-0.07}$ & $2.10^{+0.10}_{-0.19}$ \\
$H_\alpha$ & $6563\AA$ & $0.149^{+0.016}_{-0.010}$ & $0.309^{+0.027}_{-0.04}$ & $0.71^{+0.05}_{-0.08}$ \\
\enddata  
\label{tab:pearson}
\tablecomments{Pearson correlation coefficients ($\rho$) and log-linear fit parameters ($m,\,b$) between FUV and NUV emission lines and the mass accretion rate. These coefficients include all 25 epochs of HST observations, and are calculated in $\log_{10}$ space. While $m$ and $b$ are unitless, the line luminosities and mass accretion rates for these log-linear relations are written in terms of $L_{line}/L_\star$ and $\dot{M}/1\times10^{-8} M_\odot \, yr^{-1}$ (see Eqn. 5).}
\end{deluxetable}

$H_\alpha$ in particular is useful as an accretion tracer because it is bright and easily observable from the ground. We found a limited correlation between $H_\alpha$ and $\dot{M}$, with a $\rho$ of 0.149, which is in contrast to other studies \citep[e.g.,][]{white03, natta04, ingleby13}. However, we note that a strong correlation would be readily apparent if the points belonging to TW Hya in the lower right corner of the final panel in Fig. \ref{fig:big_corr} were not included, with a $\rho$ of $.780^{+0.013}_{-0.011}$ and a slope and offset of $1.25^{+0.03}_{-0.05}$ and $2.67^{+0.08}_{-0.18}$ respectively. TW Hya is in a pole-on viewing geometry \citep{andrews16}. Modeling suggests that sources with low inclinations should have stronger $H_\alpha$ emission \citep{lima10}, which is consistent with our findings. In general, TW Hya appears to be have stronger line emission than the trends shown in other objects, suggesting it is an outlier. Lines where TW Hya is less apparent as an outlier include $Si_{II}$, $S_{III}$, $Al_{III}$.
Measured $H_\alpha$ line luminosities are also affected by changes in wind opacity along the line of sight which can appear as variable blue-red line asymmetries. 
This effect likely increases the scatter in the measured relationship between accretion luminosity and $H_\alpha$. 

One notable spectral feature is a broad continuum bump centered near $1600 \si{\angstrom}$ with a width of $\sim30\si{\angstrom}$ produced by $H_2$ \cite{bergin04,france17}. A planned companion paper will address this feature within the scope of this sample in more depth.

\subsubsection{U-band excess as an accretion diagnostic}
\begin{figure}
    \centering
    \includegraphics[width = .95\linewidth]{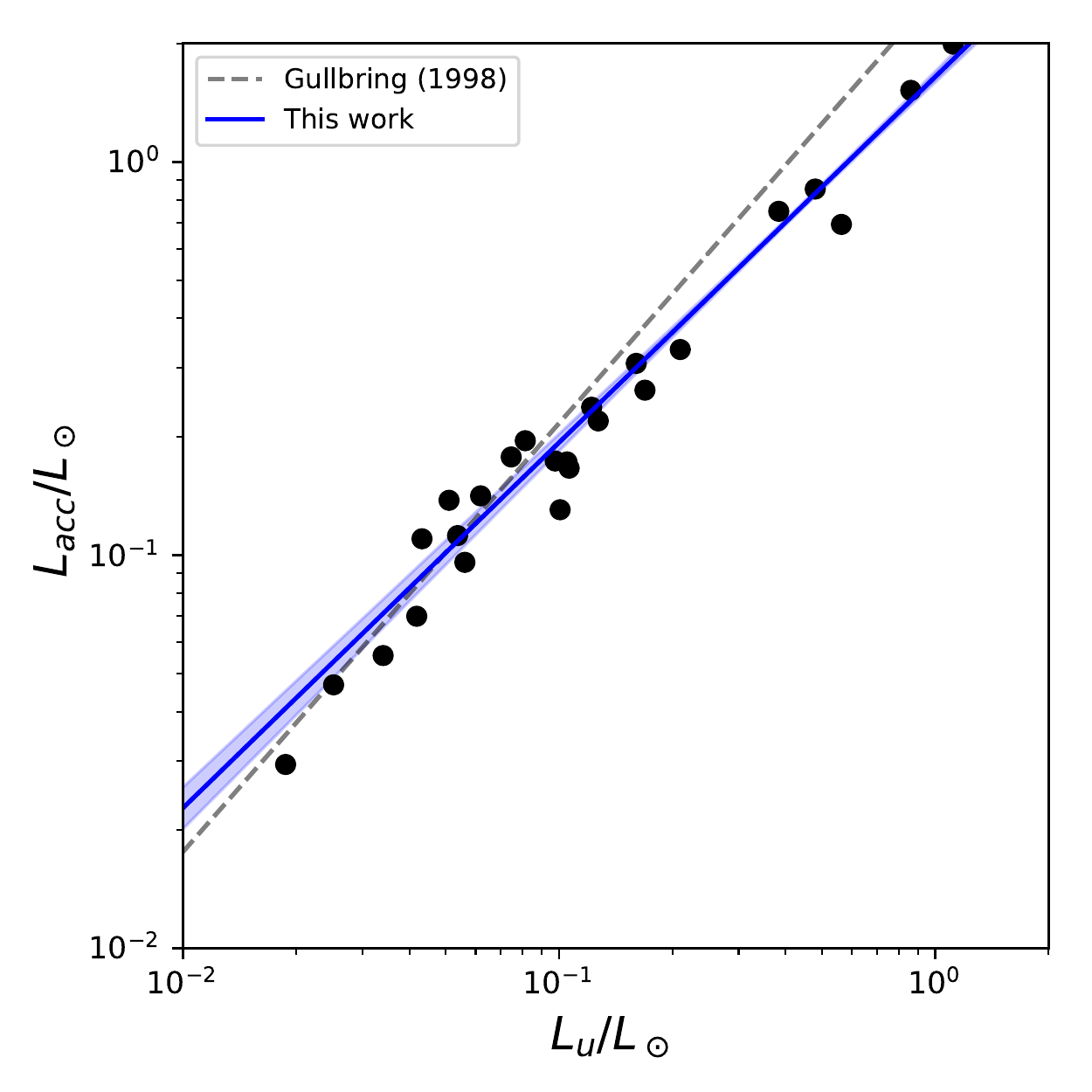}
    \caption{Accretion luminosity as a function of excess U-band luminosity. The relation found from our analysis is shown in blue (see Eqn. 7), and the commonly used relation from \citet{gullbring98} is shown by a broken black line}.
    \label{fig:Lacc}
\end{figure}

Empirical relations between excess flux in the Johnson U-band \citep{johnson53} and the mass accretion rate are commonly used to infer mass accretion rates from the ground. Our sample of 25 observations provide an opportunity to test and revise these relations. After de-reddening, we subtracted the non-accreting photospheric template from our stars and convolved the spectrum of the excess in the optical with the Johnson U-band filter. Next, accretion continuum luminosities were found from our shock models by assuming 
\begin{equation}
    L_{acc} = \frac{GM_\star\dot{M}}{R_\star}\Big(1 - \frac{R_\star}{R_{in}}\Big).
\end{equation}
We find the following linear relation for converting from excess U-band luminosity to total accretion luminosity
\begin{equation}
    \ln(L_{acc}/L_\odot) = 0.93^{+0.03}_{-0.03} \ln (L_U/L_\odot) +0.50^{+0.03}_{-0.03}.
\end{equation}
We then compare our results to the commonly used empirical relation for measuring accretion rates from \citet{gullbring98}. These two relations and our accretion luminosity measurements as a function of convolved U-band flux are shown in Fig. \ref{fig:Lacc}. We find that our results and those of \citet{gullbring98} are in relatively good agreement given the complexities of this type of analysis. An identical analysis was performed using the Sloan u' filter response function and recovered a slope of $0.91^{+0.03}_{-0.03}$ and an offset of $ 0.42^{+0.03}_{-0.11}$. Note that these relationships are written using the natural log for comparison against the \citet{gullbring98} relation (unlike the relationships in the previous section). 

One major issue with using this type of method to obtain mass accretion rates is that it is highly dependent on the ability to accurately remove the photospheric contribution. Here we use the posterior of the photospheric scaling factor $s$ from our analysis, but clearly that will not be available for most objects. Alternatives include more conventional measurements of photospheric line veiling, or photosphere fitting from longer wavelength observations that are less veiled by the accretion shock. Using longer wavelength photometric observations (i.e., NIR) comes with its own risks, namely the possibility of veiling from the protoplanetary disk rather than the accretion shock. Accurate measurements of the extinction are also important for any determination of the accretion rate relying on continuum excess.

\section{DISCUSSION}

We have presented measurements of the accretion rate and surface coverage of accretion columns of the stellar surface for a sample of five young stars with multiple contemporaneous FUV to NIR observations. In addition, we have measured contemporaneous line luminosities for several spectral features in the NUV and FUV which have previously been used as diagnostics for accretion. Here we interpret our results, examine correlations, and discuss interesting epochs. 

\subsection{Overall Variability}
We found significant amounts of mass accretion variability in all of the objects, even on the shortest timescales included in this sample. Even the objects that showed the least amount of variance still changed by $50\%$ on timescales of a week, and showed large changes in FUV - NIR continuum and line fluxes. In addition to large changes in the mass accretion rate, significant changes in how mass is distributed between columns of different densities were observed on short timescales. As indicated by previous studies of this nature \citep{ingleby15}, these rapid timescales suggest that variability is induced by changes in the very inner parts of the disk since the viscous timescale is on the order of years for a standard disk model near 1 AU \citep{hartmann98}. Whether this variability is induced by a changing magnetosphere or inhomogeneities or the inner disk structure or perhaps more likely a combination of both is more difficult to ascertain from these observations alone. Observed variability is further complicated by rotational modulation from hot spots and cool star spots rotating in and out of view. Spectropolarmetric observations of these objects would greatly help place the observed UV variability in the context of magnetic field structure. 

\subsection{Remarks on Individual Objects}
In the following sections we discuss each object in depth, focusing on the mass accretion rate and surface coverage by columns of different densities. 
When possible, we attempt to place our observations in context with broad photometric surveys of variable young stars by providing suggestions as to the nature of the objects in our sample based on the classification schemes from \citet{cody14}. Within the sample presented here, we identify GM Aur as a candidate burster and SZ 45 as likely either a quasi-periodic or stochastic source. TW Hya has already been identified as being dominated by flicker noise with semiperiodic changes attributed to instabilities in the inner disk \citep{rucinski08, siwak11a, siwak11b, siwak14, siwak18}. Our results for TW Hya are consistent with those findings. Other sources are harder to identify due to limited number of observations (DM Tau) or irregular signatures (VW Cha). 
Although some of the differences between our measurements and previous measurements of the mass accretion rate can be attributed to systematic differences between accretion diagnostics, we do find large changes between our own observations from epoch to epoch. We also find that the mass accretion rates presented here can be used to derive similar relationships to previous diagnostics (see Fig. \ref{fig:Lacc}). This suggests that the primary driver in differences between our inferred mass accretion rates and previous works is likely changes in the overall mass accretion rate and how much of the accretion excess is visible due to rotational effects.

\subsubsection{DM Tau}
DM Tau is a M1 star \citep{ingleby13} in the Taurus-Auriga region with an inner disk cavity \citep{andrews11}.
DM Tau has previously been modeled using the \citet{calvet98} accretion shock models, finding a value of $2.9\times10^{-9}\, M_\odot \, yr^{-1}$ \citep{ingleby13}.
An earlier accretion measurement by \citet{valenti93} found a similar accretion rate. 

The mass accretion rate for DM Tau ranges from $\sim2.0 - 3.6 \times10^{-8} M_{\odot} \, yr^{-1}$ over three observational epochs. Epoch 1 is separated from Epoch 2 by one week and Epoch 3 was obtained roughly 3 months later. DM Tau is an example of a transitional disk with a significant amount of accretion. The change in mass accretion rate observed in DM Tau between Epochs 1 and 2 is comparable to much of the variability observed in this sample of full/pre-transitional disks and GM Aur aside the outburst during Epoch 7 (see \S 5.2.2).

One unique feature about DM Tau is the high fraction of mass accreted via high density columns. Over $60\%$ of the mass/energy deposited onto DM Tau is carried by the high density columns for all observational epochs, with a maximum of $91\%$ during Epoch 3. We speculate that this might be a signature of a magnetic field with strong higher order multipole (e.g., octupole) components. Analytic work and simulations predict that magnetically controlled accretion flows with higher order multipole components should have higher energy fluxes at the surface of the star with smaller surface coverages than dipole controlled flows \citep{adams12, robinson17}. 
At the shortest wavelengths included in our analysis, we find some degree of excess that cannot be explained with the three component model. A discussion of four column fits with a higher column energy flux is presented in \S 8.2. The results from this additional analysis are consistent with the interpretation presented here, but with slightly higher mass accretion rates and higher degrees of surface coverage by low density shocks. 

\subsubsection{GM Aur}
GM Aur is a well studied system containing a transitional disk around a young K5 \citep{manara14} solar analogue star in the Taurus-Auriga complex. SED modeling and sub-mm observations of GM Aur show the presence of a large ($\sim 35$ AU) cavity with some remnant optically thin dust near the star \citep{calvet05, hughes09, andrews11, espaillat11, macias18}.

Measured accretion rates for the transitional disk GM Aur range from $\sim0.6\times10^{-8}$ to $2.0 \times 10^{-8} M_\odot \, yr^{-1}$ with significant changes in accretion over week and month long timescales. 
The first 3 epochs of observations presented in this paper have previously been modeled using the \citet{calvet98} shock models \citep{ingleby15} and measured accretion rates of $1.1\times10^{-8}, \, 8.5\times10^{-9},$ and $3.9\times10^{-9} \, M_\odot \, yr^{-1}$.
We find higher degrees of surface coverage by low density columns ($1\times10^{10} ergs \, s^{-1} \, cm^{-2}$) and higher values of accretion rate. These differences are likely primarily due to differences in how photospheric scaling was treated in each respective analysis, in addition to the other updates to the model discussed previously in \S 3. We find higher degrees of veiling than the constant value of $r_V = 0.2$ adopted by \citet{ingleby15}. Because the shape of the emission from low density columns most closely resembles photospheric emission, it seems reasonable that we would measure higher contributions in this work given our different fitting procedure. Surface coverage for the medium and high density accretion columns remains largely unchanged in both analyses.
The accretion behavior of GM Aur during Epochs 4 and 5, taken 4 years later than Epoch 3, appear fairly consistent with observations taken in Epoch 1 and 2. 

One of the more striking features from our analysis are the changes in emission observed over Epochs 6, 7 and 8 of the GM Aur observations (see Fig. \ref{fig:allspectra}). These data were taken over two weeks with roughly a week separation between each observation. Continuum levels in the FUV are higher in Epoch 7 than the other epochs by roughly a factor of 3, while emission in the NUV during this epoch is higher than previous epochs by between a factor of 3 and 10 (depending on the epoch/wavelength). 
The inferred accretion rate increased from $\sim 0.6$ to $2.0 \times10^{-8} \, M_\odot /yr$ between Epoch 6 and 7, and then returned to $\sim1.0 \times10^{-8} \, M_\odot /yr$ during Epoch 8. 

An upper limit set by the observational cadence can be placed on the relevant timescale for this event of about 2 weeks. The rotation rate of GM Aur has been measured to be 6.1 days \citep{percy10} which is roughly the same as our cadence for these three epochs. Our interpretation is that we are seeing roughly the same region of the star in each of our observations. This implies that the changes in the spectrum are due to changes in the mass accretion rate rather than stable hot spots rotating in and out of view. 

Our analysis for Epoch 7 indicates that roughly 30\% of the surface of the star is covered in low density columns, which is the highest ever measured for this object. Our analysis shows that $\sim0.2\%$ of the star is covered by high density accretion columns, which is a full order of magnitude higher than Epoch 6 and roughly triples the next highest coverage ever measured in all epochs. Epoch 7 also has the largest percentage ($43\%$) of mass/energy contributed by the high density columns while most of the other epochs have contributions of $10-20\%$. Epoch 8 has the second highest contribution at $27\%$, perhaps consistent with residual higher density material from the event that occurred during Epoch 7. Discussion and fits for a model of Epoch 7 with contributions from a fourth column with an even higher energy flux of $3\times10^{12} \, ergs \, s^{1} \, cm^{-2}$ is included in the Appendix in \S 8.2. The inclusion of this higher density column does not significantly change the interpretation presented here. 

GM Aur appears to satisfy the trends that would classify it as a burster as described by \citet{cody14}. This type of object is categorized by rapid increases in the accretion rate followed by a slower decay with moderate degrees of veiling and face-on inclinations \citep[see][]{stauffer14}. GM Aur has a measured inclination of $52.77^\circ {^{+0.05}_{-0.04}}$ from radio interferometry observations \citep{macias18} which is consistent with the picture of bursters. 

\subsubsection{SZ 45}
SZ 45 (alternative name: T56) is a system in Chamaeleon comprised of a M0.5 star \citep{manara14} surrounded by a pre-transitional disk \citep[][]{kim09}. Detailed spectral energy distribution (SED) modeling of the IR excess estimates the extent of the gap between the inner and outer disk to be 20AU \citep{espaillat11}. Previous measurements of $\dot{M}$ found $5\times10^{-9} \, M_\odot \, yr^{-1}$ \citep{manara14}.

The mass accretion rate from our measurements for SZ 45 varies between $\sim0.9 - 1.6 \times10^{-8} M_\odot \, yr^{-1}$ during the five epochs that it was observed. During the first four epochs, which were taken over the course of 1 week, the mass accretion rate steadily increases up to roughly a factor of 2 higher than the first observation. During the fifth and final observation (1.5 months later), $\dot{M}$ returned to a similar value as Epoch 2 of $\sim 1.2\times10^{-8} \, M_\odot \, yr^{-1}$.

One possible explanation for the slow consistent rise of the mass accretion rate over timescales of days is a stable hot spot rotating into view.
Rotation rates of CTTS and WTTS from large photometric surveys reveal a distribution of rotation periods centered around 5.2 days \citep{venuti17}, which is comparable to the total amount of time covered by the first four epochs of observations. A large fraction of the variability in young stars has been found to be simply due to rotational effects \citep{venuti15}.
SZ 45 on average has moderate surface coverage during all of the observations, ranging from $17-30\%$ surface coverage. 
Without magnetic field topography readily available for this object, one might speculate that this smooth increase in accretion rate and surface coverage is a signature of an ordered field perhaps with dipolar field components. 
A multipole field with predominately dipole contributions is thought to lead to hot spots concentrated towards the poles with limited surface coverage \citep[e.g., TW Hya;][]{donati11}. Additionally, an object with a strong dipole field should have smaller contributions from high density columns compared to an object with a strong higher order field due to the decreased amount of compression of the flow near the star \citep{adams12, robinson17}, which is observed in this object, with typical energy contributions from low density columns comprising roughly $70\%$ of the total energy budget.

One discrepancy in this interpretation is that our analysis suggests that the ratio of mass being transferred by the medium and high column densities is changing throughout these epochs. These changes perhaps indicate that instead of solely rotationally induced changes, SZ 45 is closer to the quasiperiodic variability class described by \citep{cody14} or possibly the stochastic object class. Both classifications exhibit some degree of stochastic variability over these timescales analogous to what was observed.

Longer term photometric monitoring of this system would be helpful for further interpreting these results.

\subsubsection{TW Hya}
TW Hya is one of the closest examples of a protoplanetary disk at a distance of 60 pc \citep{gaia16, gaia18b}. It is a transitional disk system in the TW Hya association with a K7 central star \citep{manara14}. TW Hya has been studied extensively due to its proximity and relative isolation \citep[e.g.,][]{calvet02, andrews16}. Accretion rates from UV/optical spectra for TW Hya have been measured before with values of $2.3\times10^{-9} M_\odot \, yr^{-1}$ \citep{ingleby13} and $1.3\times10^{-9} M_\odot \, yr^{-1}$ \citep{manara14}. Other measurements of the mass accretion rate from secondary indicators give similar values of $\sim 1\times10^{-9} M_\odot \, yr^{-1}$ \citep[e.g.,][]{donati11, curran11}.

The inferred mass accretion rates across the four epochs of observations of TW Hya were consistently the lowest in this sample, with values ranging between $\sim1.4$ to $3.3\times10^{-9} M_\odot \, yr^{-1}$, which are roughly consistent with previous measurements. Changes in the accretion rate on the order of $\sim2$ were observed across all of the epochs, which is fairly typical within this sample. Separations between our observations range from $\sim1$ week, 3 months and 5 years. 
The total surface coverage of accretion columns varies from $~9-29\%$.
TW Hya has been observed to be nearly face on \citep[$i = 7^\circ \pm 3^\circ$][]{andrews16}, which should limit variability produced by hot spots or star spots rotating in and out of view. 

No single column density greatly dictates the influx of material/energy onto TW Hya. There are moderate changes in the percentage of mass contributed by each component. Epochs 1, 2 and 3 which were all taken within a few months of each other have roughly equal relative contributions to the total mass influx from the high density columns ($20\%$ to $25\%$), while the remainder varies greatly between the medium and low density columns. Epoch 4, taken 5 years later, has has a slightly higher contribution from high density columns at $34\%$. 

TW Hya is the only object in our sample that has detailed mapping of the magnetic field topography through spectropolarimetric  observations \citep{donati11}. Those observations indicate the presence of a predominantly poloidal field dominated by a rotation axis aligned octopolar component with a weaker antiparallel rotation axis aligned dipolar component. Those authors found accretion powered hot spots with coverage on the order of $2\%$, which is similar to the coverage by medium and high density columns indicated by our analysis. It was also found that accretion occurred primarily at the poles of the star, but there was some evidence of variable accretion at lower latitudes, possibly consistent with the modest surface coverage of low density columns that we have found with our analysis. Long term photometric surveys of TW Hya indicate flicker noise behavior and longer term quasiperiodic behavior, which is consistent with the type of variabilty observed during these four epochs \citep{rucinski08, siwak11a, siwak11b, siwak14, siwak18}. 

\subsubsection{VW Cha}
VW Cha is system that includes a full disk around a K7 star located in Chamaeleon \citep{manoj11}. Previous observations of VW Cha have measured mass accretion rates of $1.1\times 10^{-7} M_\odot \, yr^{-1}$ \citep{hartmann98}, $2.5 \times 10^{-8} M_\odot \, yr^{-1}$ \citep{manara16a} and $3.2\times10^{-8} M_\odot \, yr^{-1}$ \citep{salyk13}. 

Our analysis for VW Cha reveals a robust accretor with significant contributions from both low and high density accretion columns. VW Cha is the strongest accretor in our sample, with median mass accretion rates ranging between $\sim6.91 - 19.9 \times10^{-8} M_{\odot} \, yr^{-1}$ over the five epochs that it was observed. The first 4 epochs were observed over roughly 1 week and Epoch 5 was obtained 1.5 months later. 

While all of the epochs display variability in the FUV and NUV typical of all of the objects in our sample, the NIR and optical wavelengths show interesting behavior. The first 4 epochs of VW Cha are roughly consistent with each other in the optical and NIR, but we find a large decrease in both wavelength ranges during Epoch 5 (see Fig. \ref{fig:allspectra}). The FUV and NUV emission during Epoch 5 is higher than 2 out of 4 of the previous epochs. This is inconsistent with the typical variability observed in other objects in this sample.
To explain this curious behavior, we consider several possibilities: (1) pointing error, (2) contamination from a companion, (3) star spots, (4) variable global extinction, and (5) a partial chance occultation by the magnetosphere or a disk warp along the line of sight.

\paragraph{(1) Pointing Error}
We inspected the centering of the object on the slit from the \textit{HST} acquistion images, and VW Cha appears to be well centered in the slit. Aperture photometry of VW Cha using the acquisition images in the F28X50OII observing mode (centered on $O_{II}$ at $3727 \si{\angstrom}$) is consistent with the flux changes observed in the spectrum, suggesting that the object was indeed well centered in the slit. 

\paragraph{(2) Companions}
VW Cha is a known triple system \citep{brandeker01}. The primary and secondary stars are separated by 0.\arcsec7, and the tertiary is tightly bound to the second star with a projected separation of 0.\arcsec1. The secondary star is well resolved and readily apparent in the acquisition images. The tertiary star remains undetected by eye due to its dimness and proximity to the secondary. During Epochs 1,2,3 and 4, the position angle of the slit is set such that only the primary star fell onto the slit. During Epoch 5, the position angle of the slit changed significantly causing the secondary to also align with the slit, which is visible in the 2D spectra. The stars are far enough apart on the detector to avoid significant contamination from the secondary during extraction, and care was taken to ensure that the background level selected for subtraction is also far away from either star. The background emission is faint and relatively flat across the detector.
In addition, if there is contamination from the secondary in the extraction of the primary it would instead act to increase the flux which is opposite of what was observed. All of this suggests that contamination from the secondary (or the dim tertiary) is not responsible for the changes in flux. Additional companions falling onto the slit due to changes in the position angle are ruled out by high resolution (0.1\arcsec) images of VW Cha by \citep{brandeker01}.

\paragraph{(3) Star Spots}
Typical star spots are $\sim 500K$ cooler than the photosphere \citep{venuti15}, and will decrease the emission in the optical and NIR. Both the star spots and the undisturbed photosphere contribute little to the observed UV excess compared to the contributions from the accretion shock \citep[see][]{calvet98, ingleby13}. Since we have made the assumption that the undisturbed photosphere is not changing between different epochs, this is not something that our model can recover. Lifetimes of star spots on TTS are on the order of years \citep{bradshaw14}. Given this, it seems unlikely that changes of this magnitude would occur in spot coverage over a period of 1.5 months, making this a less plausible explanation for the observed changes. A large spot hidden on the opposite side of the star during the first four epochs is also an unlikely explanation for the observed changes. Epochs 1-4 span roughly 1 week, which is comparable to the typical rotation rate observed in young stars \citep[see ][]{bouvier07}. If there was a large cool spot, its effects likely would have been observed during those epochs.

\paragraph{(4) Changes in Global Extinction}
Changing extinction column densities is a common source of variability in young stars \citep[e.g., AA Tau][]{grankin07, bouvier13}. 
The expected signature of an increase in extinction is a large decrease in the observed FUV and NUV, with a smaller decrease in the optical in NIR. This is inconsistent as the sole cause of the observed changes in Epoch 5 because the FUV and NUV are similar to previous epochs while the optical decreases and we would expect to see the largest changes in the UV. 

\paragraph{(5) Partial Occultation}

The situation changes if only part of the star is extincted. If the degree of accretion is similar to other epochs, one could imagine a scenario where a chance alignment of the cooler magnetosphere or an inner disk warp covers a large fraction of the undisturbed photosphere, but not the regions of the star covered by high energy flux accretion columns. This would result in suppressed photospheric emission in the optical/IR while the emission in the NUV/FUV would remain primarily unchanged, which is consistent with our observations.
The amount of material blocking stellar emission along the line of sight is difficult to directly measure due to degeneracies between accretion, the fraction of the star that is occulted, and occulting column density. However, if we make the simplifying assumption that the extincting screen is fully opaque to the stellar emission, we can estimate the fraction of the star that is blocked and obtain lower limits on the mass of the screen. To make this approximation, ${f_i}$, $w$, and $s$ for Epochs 1 - 4 were re-fit simultaneously, while the parameters for Epoch 5 were fit separately. Assuming all of the changes in the photospheric scaling factor $s$ for Epoch 5 are solely due to occultation, we find that approximately $60\%$ of the emission from the stellar surface is blocked. Using the typical relation between $A_v$ and column density for $R_V = 3.1$ ISM dust \citep{weingartner01} and a uniform extinction across the screen of 3 $A_v$ (such that $\sim~95\%$ of the light is absorbed at V band), we find a very rough estimate of the minimum amount of occulting dust of $M_{occult} > 1\times10^{-14} \, M_\odot$. 

The fraction of coverage observed in VW Cha is comparable to the inferred coverage of the young star V354 Mon, where disk warps were invoked to explain changes in observed flux \citep{fonseca14, schneider18}. One complication in this explanation for VW Cha is that V354 Mon is extincted by the disk warp about half of the time with a period of 5.25 days, where we only observe these changes in one out of five epochs. The first four epochs span roughly 1 week, which is similar to typical rotation rates of CTTS. To avoid this issue, one might speculate that the disk warp has a longer period than 7 days (giving a Keplerian radius of $ > \sim0.08$AU.)

Extinction by the magnetosphere requires a moderate amount of dust grains within the accretion column to sufficiently extinct the light from the photosphere.  Previous observations have suggested that large grains may be able to survive near the co-rotation region \citep[e.g.,][]{ingleby15, stauffer15}. This would suggest that the canonical picture of a dust-free inner gas disk may be over simplified. If occultation by a dusty column is more sensitive to geometric alignment than disk warps, it might explain why we do not see its effects during the previous four epochs as the duration of the occultation would be reduced.

While extinction by the accretion column is our preferred explanation, we note that the above is qualitative and speculative since our models do not recover this sort of geometric effect and we cannot discern between different alignment scenarios under the current framework.  In any case, the changes in the observed flux appear to be astrophysical in nature, and not a by-product of calibration. Given the strange behavior observed for this object, it is difficult to speculate with any certainty on a variability class, but if the observed changes are indeed due to extinction, then it may fall into the ``dipper'' class. Additional observations, particularly photometric monitoring to better constrain occultation timescales and sub-mm interferometric observations to measure the inclination of the disk, would greatly help with our interpretation of these observations.

\subsection{Star spots and Caveats}
Previous efforts with similar modeling processes have scaled the photospheric flux by a constant veiling value generally measured from a high-resolution often nonsynchronously obtained optical spectrum. This approach results in undisturbed photospheric fluxes that are inconsistent if variability in the accretion shock emission occurs. 
The analysis presented here makes the assumption that the continuum fluxes from the undisturbed photosphere for an individual star remain constant between observations. In reality, the emission from the photosphere will likely be modulated by star spots. Surface coverage from stars spots can be large ($\sim20\%$) for young stars, but is typically very long lived \citep{bradshaw14}. While changes in mass accretion rates may happen on shorter timescales than changes in star spot distributions, rotational modulation of existing star spots occurs on similar timescales.

Disk warps are another possible source of variability that were not considered in this analysis for objects other than VW Cha. If the disk is not symmetric, raised regions of the disk can pass between the star and the observing manifesting as changes in the extinction as a function of disk rotational phase \citep[e.g., AA Tau][]{grankin07, bouvier13}. This effect primarily  occurs if the inclination is such that the disk is being view nearly edge-on. It is difficult to infer the importance of disk warps with the limited cadence of our \textit{HST} observations, but additional photometric monitoring or mm observations to obtain inclinations for these objects would make it possible to identify cases where disk warps are an important component of the variability.

Another important caveat to the work done here to keep in mind is that stars are inherently 3D objects, and we are inferring the accretion rates and surface coverage based only on the excess visible to us as the observer. Based on the paradigm of magnetospheric accretion and the localization of magnetic footprints shown by spectropolarimetric measurements \citep[e.g.,][]{donati11b} and hot spot modulated light curves, it is very possible that surface coverage on the far side of the star is different than surface coverage on the front side. Likewise, the filling factors provide only a crude method of adjusting the amount of excess produced by the model compared to a a full 3D model with geometric effects as discussed under the context of VW Cha in \S5.2.5.  

With this in mind, some of the variability observed here can likely be explained by changes in which parts of the star are visible to us instead of changes in the amount of material falling onto the star. Higher cadence observations and measurements of the rotation rate for individual stars would help eliminate some of these concerns and aid in the interpretation of these spectra.

\section{CONCLUSIONS}
Using multiple \textit{HST} UV spectra of individual young stars, we have explored accretion variability from three approaches: the overall mass accretion rate, coverage of the different density accretion columns, and FUV/NUV emission lines. Here we summarize the main points of this work.

\begin{enumerate}
    \item We have updated the \citet{calvet98} shock models and include veiling and model uncertainty as parameters when fitting multi-epoch observations.
    \item We find that mass accretion rates are highly variable, with typical spreads of roughly an order of magnitude for an individual object. As such, future studies considering an effect that may be sensitive to FUV/NUV emission \citep[e.g., infrared $H_2$ line emission;][]{nomura07} should consider obtaining a simultaneous diagnostic of the accretion rate.
    \item U-band photometry and FUV, NUV, and $H_\alpha$ line luminosities have been found to be indicators of mass accretion rates, which is consistent with previous findings.
    \item The transitional disk GM Aur underwent a factor of 3.5 increase in the mass accretion rate in one epoch of our observations, which we interpret as an accretion burst. 
    \item VW Cha demonstrated interesting behavior during Epoch 5 in which the optical significantly decreased while the NUV/FUV remained comparable to previous epochs. Our qualitative interpretation that we offer to explain this event is a chance alignment of the accretion column leading to selective extinction of the undisturbed photosphere along our line of sight while high energy shocks remain free from extinction. 
    \item We offer suggestions as to the morphological variability classes of GM Aur as a burster and SZ 45 as either a quasiperiodic or stochastic source. The observed behavior of TW Hya is consistent with the previous interpretations of a pole-on quasiperiodic source punctuated by short-timescale flicker noise. DM Tau and VW Cha are more difficult to classify due to the limited number of epochs and unique behavior, respectively. 

\end{enumerate}

Our results continue to demonstrate the near-ubiquitous nature of variability in T Tauri stars, and highlight the need for continued UV monitoring of accreting young stars.

\section{ACKNOWLEDGEMENTS}

We would like to thank Dr. Nuria Calvet and Dr. Laura Ingleby for sharing their knowledge and insights on accretion shock modeling. We thank Dr. Philip Muirhead for useful discussions regarding the fitting methods employed in this publication. We also thank the anonymous reviewers whose comments have significantly improved this work.

The authors acknowledge funding support from \textit{HST} grant No. GO-14048, GO-14193, and GO-15165.

Based on observations made with the NASA/ESA \textit{Hubble Space Telescope}, obtained from the Data Archive at the Space Telescope Science Institute, which is operated by the Association of Universities for Research in Astronomy, Inc., under NASA contract NAS 5-26555. These observations are associated with programs 11608, 11616, 13775, 14048, 14193, and 15165.

This work has made use of data from the European Space Agency (ESA) mission {\it Gaia} (\url{https://www.cosmos.esa.int/gaia}), processed by the {\it Gaia} Data Processing and Analysis Consortium (DPAC, \url{https://www.cosmos.esa.int/web/gaia/dpac/consortium}). Funding for the DPAC has been provided by national institutions, in particular the institutions participating in the {\it Gaia} Multilateral Agreement.

\section{Appendix}

\subsection{Model fits for each epoch}
The fits for each epoch are shown in Fig. \ref{fig:fits}. The median for each component of the model is shown, along with the total model in red. 
The undisturbed photospheric contribution is also shown. It is important to keep in mind that the undisturbed photospheric contribution may be different between epochs since it is also scaled by the fraction of the star not covered by accretion columns. The unscaled version of the photosphere remains the same between all epochs of the same object, due to our treatment of the photospheric scaling factor as a single parameter which is described in depth in the \S 3.4.

\begin{figure*}
    \centering
    \includegraphics[width = 0.9\linewidth]{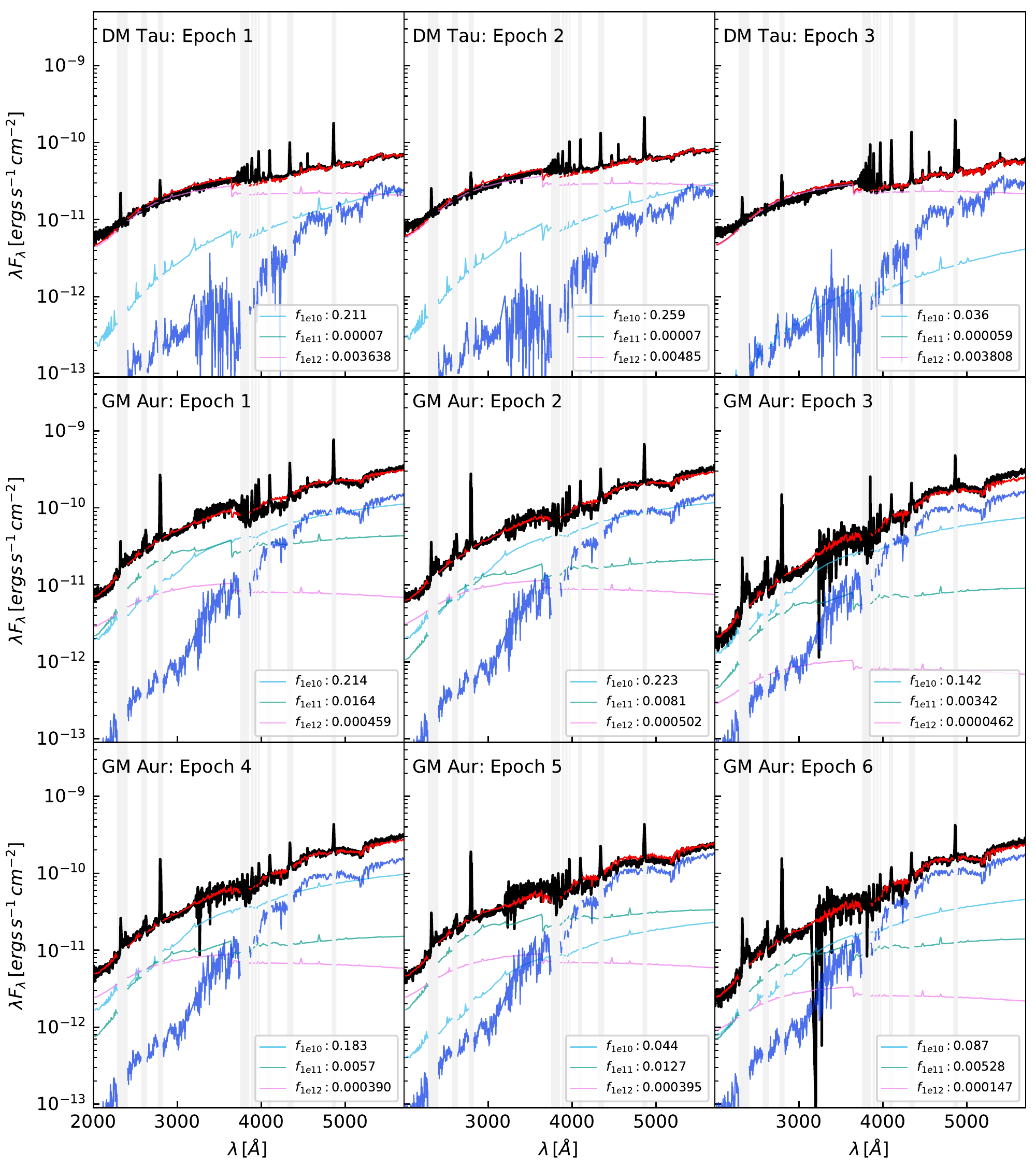}
    \caption{Model fits showing the median model. The total model is shown in red, and the contribution from the undisturbed photosphere is shown in blue. Contributions from different column energy fluxes are shown and labeled.} 
    \label{fig:fits}
\end{figure*}
\begin{figure*}
\ContinuedFloat
    \includegraphics[width = 0.89\linewidth]{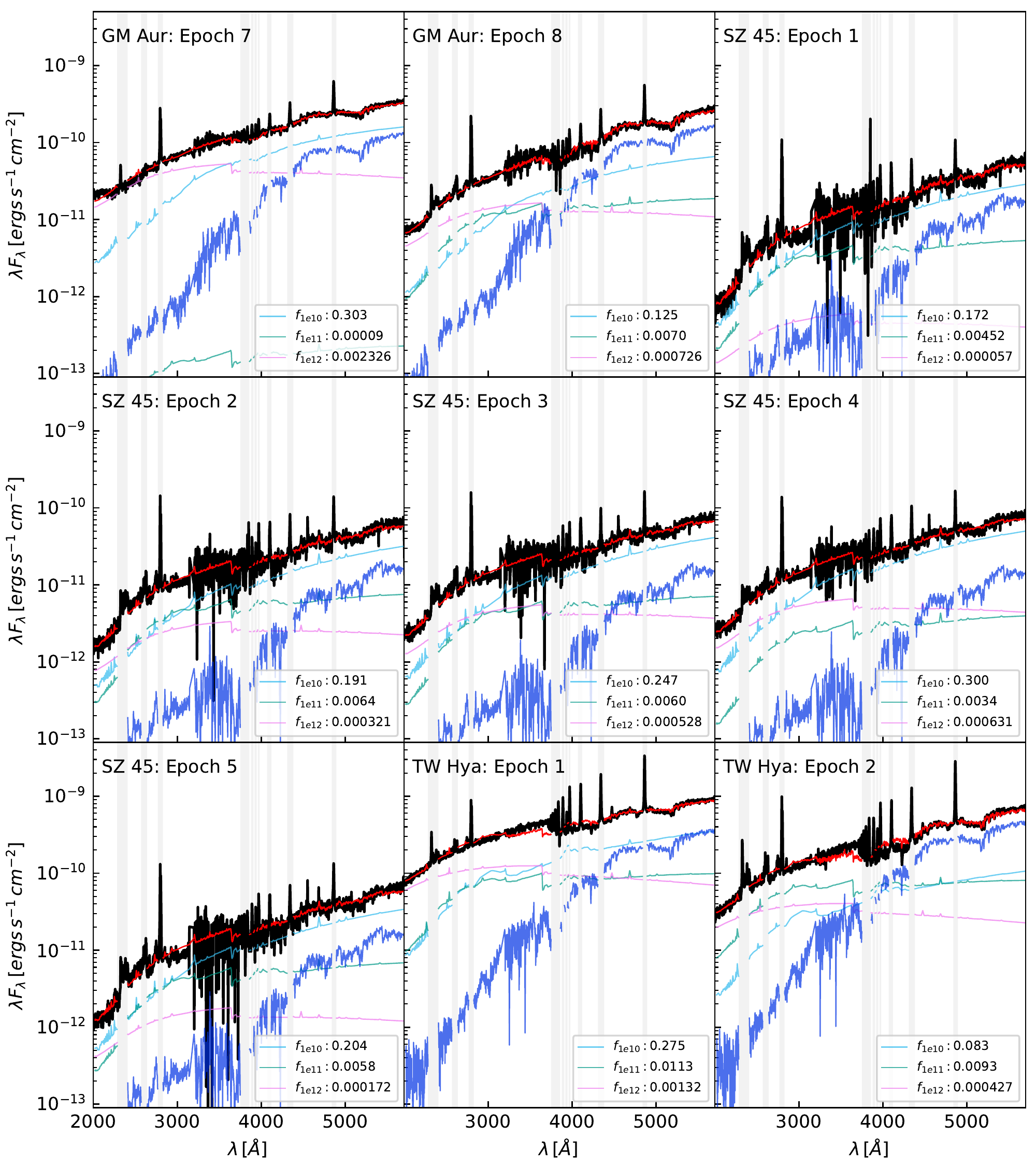}
    \caption{continued.}
\end{figure*}
\begin{figure*}\ContinuedFloat
    \includegraphics[width = 0.89\linewidth]{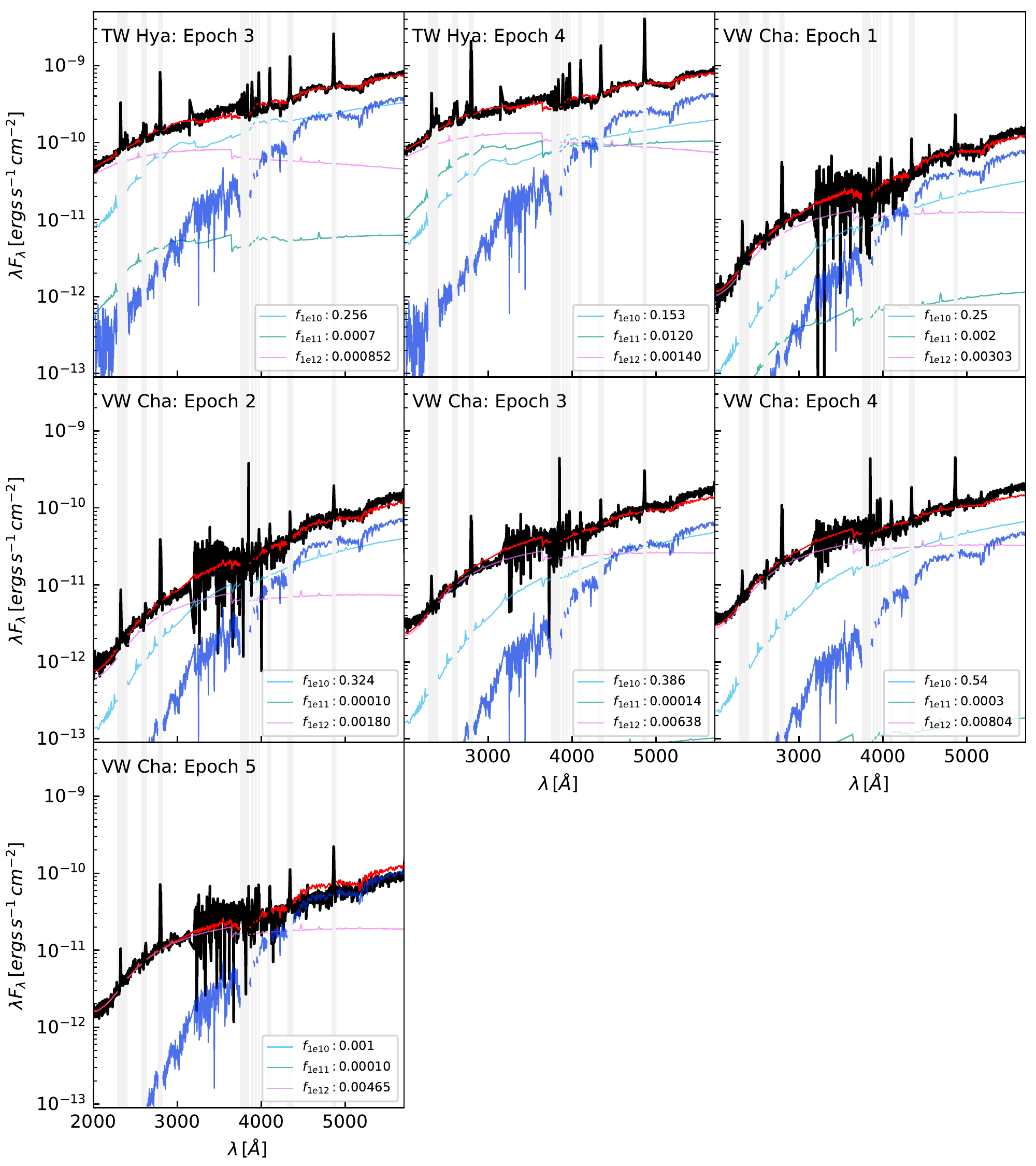}
    \caption{continued.}
\end{figure*}

\subsection{Higher energy columns for DM Tau and GM Aur}
\begin{figure*}
    \centering
    \includegraphics[width = 0.6\linewidth]{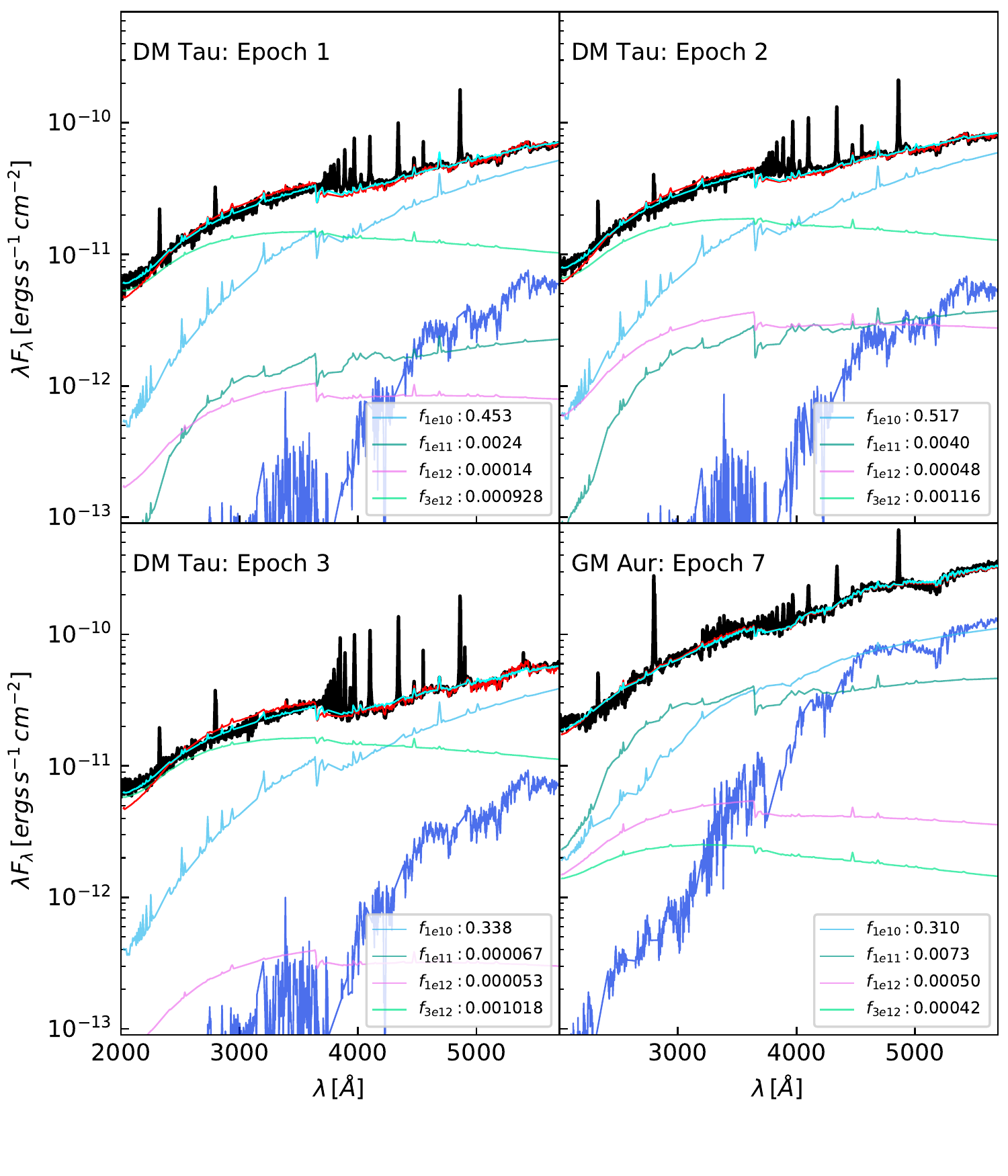}
    \caption{Models including a fourth higher column energy flux for select epochs with excess in the shortest wavelength regions. The total for the four column model is shown in cyan, while the median of the three column models is shown in red for comparison. The blue line represents the photospheric contribution and the contributions from each column are label.}
    \label{fig:four_columns}
\end{figure*}

Our fits for all three epochs of DM Tau and Epoch 7 of GM Aur using columns with $F = 10^{10}, 10^{11}$ and $10^{12} \, ergs \, s^{-1} \, cm^{2}$ are unable to reproduce some of the excess emission observed at the shortest wavelengths in the NUV. One possible explanation for this excess for DM Tau is an incorrect value of extinction. However, this does not appear to be the case, as refitting the data with $A_V = 0$ for DM Tau still resulted in a dearth of model emission in these wavelengths. Including a fourth model component with energy fluxes of $3\times10^{12} \, ergs \, s^{-1} \, cm^{-2}$ better reproduces the observed emission.
While the overall mass accretion rate does increase with the inclusion of these high density columns, this does not significantly change our interpretation presented above. The revised filling factors and mass accretion rates are shown in Table \ref{tab:fourcolumn}, and the updated fits with these components are shown in Fig. \ref{fig:four_columns}. 

Except for Epoch 7, we are able to successfully reproduce the emission in all of the other epochs for GM Aur using the three-column models. Although the change in the fit due to the addition of the higher energy column is more subtle for GM Aur Epoch 7 compared to DM Tau, the overall fit is improved and is consistent with the interpretation that a dense clump of material was accreted onto the star during this epoch which has the highest accretion rate. We find better agreement with the addition of the higher energy column between the model and data in shorter wavelength regions excluded from our fitting procedures. The inability to fit the shortest wavelength regions reiterates that this epoch is unique relative to the other 7 epochs of observations and more data would be needed to study this further. 

\begin{deluxetable*}{cccccccc}
\centering
\tablecolumns{8}
\tablecaption{Mass Accretion Rates and Filling Factors}
\tablehead{Object & Epoch & $\dot{M}$ & $f_{1E10}$ & $f_{1E11}$ & $f_{1E12}$ & $f_{3E12}$ & $r_v$}
\startdata
DM Tau & 1 & $3.707^{+0.023}_{-0.023}$ & $0.453^{+0.009}_{-0.009}$ & $0.0024^{+0.0008}_{-0.0013}$ & $0.00014^{+0.00019}_{-0.00007}$ & $0.000928^{+0.000018}_{-0.00004}$ & $5.7^{+0.5}_{-0.4}$ \\
DM Tau & 2 & $4.595^{+0.023}_{-0.024}$ & $0.517^{+0.013}_{-0.012}$ & $0.0040^{+0.0019}_{-0.0024}$ & $0.00048^{+0.0003}_{-0.00028}$ & $0.00116^{+0.00006}_{-0.00007}$ & $6.8^{+0.6}_{-0.5}$ \\
DM Tau & 3 & $3.129^{+0.019}_{-0.018}$ & $0.338^{+0.005}_{-0.005}$ & $0.000067^{+0.00004}_{-0.000016}$ & $0.000053^{+0.000014}_{-0.000006}$ & $0.001018^{+0.000007}_{-0.000007}$ & $4.8^{+0.4}_{-0.4}$ \\
GM Aur & 7 & $2.045^{+0.015}_{-0.016}$ & $0.310^{+0.008}_{-0.007}$ & $0.0073^{+0.0019}_{-0.0019}$ & $0.00050^{+0.0004}_{-0.00024}$ & $0.00042^{+0.00005}_{-0.00008}$ & $0.937^{+0.013}_{-0.012}$ \\
\enddata
\tablecomments{Mass accretion rates and filling factors for four column models for the observational epochs that showed excess at the shortest wavelengths under three column models. Mass accretion rates are presented in units of $10^{-8} M_{\astrosun} \, yr^{-1}$. }
\label{tab:fourcolumn}
\end{deluxetable*} 
\begin{rotatetable*}
\begin{deluxetable*}{r|ccccccccccccc}
\tabletypesize{\scriptsize}
\centering
\tablecolumns{15}
\tablecaption{Line Luminosities} 
\tablewidth{2.0\columnwidth} 
\tablehead{\hspace{1mm} & & & & & & & & & & & & & \\ Obj. & E. & $C_{II}$ & $C_{I}$ & $C_{IV}$ & $He_{II}$ & $O_{III}$ & $Si_{II}$ & $Si_{III]}$ & $C_{III]}$ & $C_{II]}$ & $Al_{III]}$ & $Mg_{II}$ & $H_\alpha$  \\
 \hspace{1mm}&  &$1335\si{\angstrom}$ & $1463\si{\angstrom}$ & $1548\si{\angstrom}$ & $1640\si{\angstrom}$ & $1666\si{\angstrom}$ & $1808\si{\angstrom}$ & $1892\si{\angstrom}$ & $1908\si{\angstrom}$ & $2325\si{\angstrom}$ & $2670\si{\angstrom}$ & $2796\si{\angstrom}$ & $6563\si{\angstrom}$}
\startdata
DM Tau & 1 & $18.3^{+0.8}_{-0.8}$ & $5.6^{+1.0}_{-1.0}$ & $58.3^{+2.2}_{-2.2}$ & $42.5^{+1.1}_{-1.1}$ & $1.8^{+0.7}_{-0.7}$ & $3.7^{+2.0}_{-2.0}$ & $0.5^{+0.9}_{-0.9}$ & $1.5^{+1.4}_{-1.4}$ & $11.6^{+0.9}_{-0.9}$ & $2.0^{+0.9}_{-0.9}$ & $9.3^{+1.0}_{-1.0}$ & $423.7^{+2.5}_{-2.5}$ \\
DM Tau & 2 & $16.7^{+1.0}_{-1.0}$ & $5.2^{+1.2}_{-1.2}$ & $56.1^{+2.6}_{-2.7}$ & $44.9^{+1.3}_{-1.3}$ & $1.8^{+0.9}_{-0.9}$ & $-0.0^{+2.4}_{-2.4}$ & $0.3^{+1.1}_{-1.1}$ & $0.8^{+1.8}_{-1.7}$ & $10.8^{+1.0}_{-1.0}$ & $0.3^{+0.8}_{-0.8}$ & $7.4^{+0.9}_{-0.9}$ & $480.1^{+2.5}_{-2.5}$ \\
DM Tau & 3 & $12.9^{+0.9}_{-0.9}$ & $3.2^{+1.3}_{-1.3}$ & $48.5^{+2.7}_{-2.7}$ & $36.4^{+1.3}_{-1.3}$ & $-0.2^{+1.1}_{-1.1}$ & $3.0^{+2.1}_{-2.1}$ & $1.9^{+1.1}_{-1.1}$ & $3.8^{+1.4}_{-1.4}$ & $8.9^{+0.9}_{-0.9}$ & $0.3^{+0.7}_{-0.7}$ & $12.5^{+0.8}_{-0.8}$ & $338.3^{+2.3}_{-2.3}$ \\
GM Aur & 1 & $8.97^{+0.20}_{-0.20}$ & $3.63^{+0.28}_{-0.28}$ & $42.9^{+0.5}_{-0.5}$ & $10.6^{+0.3}_{-0.3}$ & $1.18^{+0.25}_{-0.25}$ & $5.4^{+0.4}_{-0.4}$ & $1.92^{+0.22}_{-0.22}$ & $2.0^{+0.3}_{-0.3}$ & $11.5^{+0.3}_{-0.3}$ & $2.1^{+0.3}_{-0.3}$ & $76.3^{+0.5}_{-0.5}$ & $581^{+5}_{-5}$ \\
GM Aur & 2 & $8.75^{+0.22}_{-0.21}$ & $2.81^{+0.27}_{-0.27}$ & $33.3^{+0.5}_{-0.5}$ & $10.59^{+0.29}_{-0.29}$ & $0.88^{+0.23}_{-0.23}$ & $4.3^{+0.5}_{-0.5}$ & $1.56^{+0.22}_{-0.22}$ & $1.8^{+0.3}_{-0.3}$ & $10.05^{+0.28}_{-0.27}$ & $0.8^{+0.3}_{-0.3}$ & $74.8^{+0.4}_{-0.4}$ & $559^{+6}_{-6}$ \\
GM Aur & 3 & $5.52^{+0.16}_{-0.16}$ & $1.82^{+0.19}_{-0.19}$ & $16.5^{+0.4}_{-0.4}$ & $6.93^{+0.22}_{-0.22}$ & $0.50^{+0.16}_{-0.16}$ & $2.10^{+0.28}_{-0.28}$ & $0.64^{+0.15}_{-0.15}$ & $0.97^{+0.19}_{-0.19}$ & $9.11^{+0.19}_{-0.19}$ & $0.87^{+0.20}_{-0.19}$ & $43.5^{+0.3}_{-0.3}$ & $426^{+6}_{-6}$ \\
GM Aur & 4 & $5.64^{+0.17}_{-0.17}$ & $1.76^{+0.22}_{-0.22}$ & $32.0^{+0.5}_{-0.5}$ & $12.26^{+0.29}_{-0.29}$ & $0.28^{+0.19}_{-0.19}$ & $2.1^{+0.4}_{-0.4}$ & $0.67^{+0.18}_{-0.18}$ & $1.55^{+0.29}_{-0.29}$ & $8.55^{+0.27}_{-0.27}$ & $1.0^{+0.3}_{-0.3}$ & $44.4^{+0.4}_{-0.4}$ & $340^{+6}_{-6}$ \\
GM Aur & 5 & $7.76^{+0.21}_{-0.21}$ & $2.4^{+0.4}_{-0.4}$ & $37.0^{+0.6}_{-0.6}$ & $12.9^{+0.4}_{-0.4}$ & $0.4^{+0.4}_{-0.4}$ & $3.4^{+0.4}_{-0.4}$ & $1.23^{+0.19}_{-0.19}$ & $1.75^{+0.29}_{-0.29}$ & $11.03^{+0.26}_{-0.26}$ & $1.9^{+0.3}_{-0.3}$ & $62.9^{+0.4}_{-0.4}$ & $458^{+5}_{-5}$ \\
GM Aur & 6 & $5.58^{+0.17}_{-0.17}$ & $1.64^{+0.23}_{-0.23}$ & $18.3^{+0.4}_{-0.4}$ & $5.87^{+0.23}_{-0.23}$ & $0.22^{+0.19}_{-0.19}$ & $2.6^{+0.3}_{-0.3}$ & $0.51^{+0.17}_{-0.17}$ & $0.96^{+0.22}_{-0.22}$ & $10.12^{+0.22}_{-0.22}$ & $1.08^{+0.22}_{-0.22}$ & $46.0^{+0.3}_{-0.3}$ & $375^{+5}_{-5}$ \\
GM Aur & 7 & $7.6^{+0.3}_{-0.3}$ & $1.5^{+0.5}_{-0.5}$ & $44.5^{+0.8}_{-0.8}$ & $11.8^{+0.4}_{-0.4}$ & $0.9^{+0.4}_{-0.4}$ & $5.2^{+0.9}_{-0.9}$ & $3.3^{+0.5}_{-0.5}$ & $4.6^{+0.6}_{-0.6}$ & $9.6^{+0.6}_{-0.5}$ & $3.8^{+0.5}_{-0.5}$ & $75.3^{+0.6}_{-0.6}$ & $529^{+6}_{-6}$ \\
GM Aur & 8 & $7.54^{+0.20}_{-0.20}$ & $2.66^{+0.28}_{-0.28}$ & $24.9^{+0.5}_{-0.5}$ & $9.80^{+0.29}_{-0.3}$ & $0.37^{+0.24}_{-0.24}$ & $2.9^{+0.5}_{-0.5}$ & $0.68^{+0.28}_{-0.28}$ & $1.3^{+0.4}_{-0.4}$ & $9.6^{+0.3}_{-0.3}$ & $1.5^{+0.4}_{-0.4}$ & $71.3^{+0.5}_{-0.5}$ & $432^{+5}_{-5}$ \\
SZ 45 & 1 & $2.02^{+0.16}_{-0.16}$ & $1.63^{+0.21}_{-0.21}$ & $8.7^{+0.4}_{-0.4}$ & $7.00^{+0.27}_{-0.27}$ & $0.53^{+0.17}_{-0.17}$ & $2.9^{+0.3}_{-0.3}$ & $0.67^{+0.15}_{-0.15}$ & $1.11^{+0.23}_{-0.23}$ & $5.07^{+0.19}_{-0.20}$ & $1.14^{+0.21}_{-0.21}$ & $44.3^{+0.3}_{-0.3}$ & $128^{+9}_{-9}$ \\
SZ 45 & 2 & $2.32^{+0.18}_{-0.18}$ & $1.32^{+0.26}_{-0.26}$ & $11.9^{+0.5}_{-0.5}$ & $9.3^{+0.3}_{-0.3}$ & $0.45^{+0.20}_{-0.20}$ & $3.0^{+0.5}_{-0.5}$ & $0.20^{+0.22}_{-0.22}$ & $1.0^{+0.4}_{-0.4}$ & $5.5^{+0.3}_{-0.3}$ & $1.1^{+0.3}_{-0.3}$ & $57.5^{+0.4}_{-0.4}$ & $158^{+9}_{-9}$ \\
SZ 45 & 3 & $2.53^{+0.22}_{-0.22}$ & $1.7^{+0.3}_{-0.3}$ & $12.1^{+0.6}_{-0.6}$ & $10.4^{+0.4}_{-0.4}$ & $0.08^{+0.29}_{-0.29}$ & $3.8^{+0.5}_{-0.5}$ & $0.79^{+0.23}_{-0.23}$ & $1.9^{+0.4}_{-0.4}$ & $6.52^{+0.3}_{-0.29}$ & $2.9^{+0.4}_{-0.4}$ & $67.9^{+0.5}_{-0.5}$ & $191^{+12}_{-12}$ \\
SZ 45 & 4 & $2.34^{+0.20}_{-0.20}$ & $1.6^{+0.3}_{-0.3}$ & $11.4^{+0.5}_{-0.5}$ & $9.3^{+0.4}_{-0.4}$ & $0.46^{+0.27}_{-0.27}$ & $3.8^{+0.6}_{-0.6}$ & $1.29^{+0.27}_{-0.28}$ & $1.3^{+0.4}_{-0.4}$ & $5.6^{+0.3}_{-0.3}$ & $1.7^{+0.4}_{-0.4}$ & $60.6^{+0.5}_{-0.5}$ & $183^{+10}_{-10}$ \\
SZ 45 & 5 & $2.02^{+0.18}_{-0.18}$ & $1.74^{+0.25}_{-0.26}$ & $9.7^{+0.5}_{-0.5}$ & $7.3^{+0.3}_{-0.3}$ & $0.66^{+0.20}_{-0.20}$ & $4.8^{+0.4}_{-0.4}$ & $1.43^{+0.21}_{-0.21}$ & $1.36^{+0.27}_{-0.27}$ & $5.42^{+0.26}_{-0.26}$ & $0.8^{+0.3}_{-0.3}$ & $60.6^{+0.5}_{-0.5}$ & $195^{+10}_{-10}$ \\
TW Hya & 1 & $1.87^{+0.24}_{-0.24}$ & $2.2^{+0.5}_{-0.5}$ & $12.9^{+0.7}_{-0.7}$ & $7.1^{+0.5}_{-0.5}$ & $0.8^{+0.4}_{-0.4}$ & $0.11^{+0.16}_{-0.16}$ & $0.03^{+0.08}_{-0.08}$ & $0.31^{+0.11}_{-0.11}$ & $2.94^{+0.10}_{-0.10}$ & $0.02^{+0.11}_{-0.11}$ & $8.51^{+0.13}_{-0.13}$ & $162.5^{+1.7}_{-1.7}$ \\
TW Hya & 2 & $1.98^{+0.21}_{-0.20}$ & $1.5^{+0.3}_{-0.3}$ & $7.4^{+0.6}_{-0.6}$ & $4.0^{+0.4}_{-0.4}$ & $0.43^{+0.25}_{-0.25}$ & $0.61^{+0.14}_{-0.14}$ & $0.09^{+0.05}_{-0.05}$ & $0.39^{+0.09}_{-0.09}$ & $2.92^{+0.07}_{-0.07}$ & $0.02^{+0.09}_{-0.09}$ & $9.82^{+0.11}_{-0.11}$ & $164.6^{+1.6}_{-1.6}$ \\
TW Hya & 3 & $1.80^{+0.09}_{-0.09}$ & $1.12^{+0.14}_{-0.15}$ & $8.33^{+0.26}_{-0.26}$ & $4.29^{+0.16}_{-0.16}$ & $0.22^{+0.12}_{-0.12}$ & $0.35^{+0.10}_{-0.10}$ & $0.09^{+0.05}_{-0.05}$ & $0.37^{+0.07}_{-0.08}$ & $3.03^{+0.08}_{-0.08}$ & $-0.09^{+0.09}_{-0.09}$ & $8.06^{+0.11}_{-0.11}$ & $158.4^{+1.7}_{-1.7}$ \\
TW Hya & 4 & $2.93^{+0.14}_{-0.14}$ & $1.53^{+0.20}_{-0.21}$ & $8.3^{+0.4}_{-0.4}$ & $4.09^{+0.21}_{-0.21}$ & $-0.01^{+0.17}_{-0.17}$ & $0.77^{+0.12}_{-0.12}$ & $0.10^{+0.06}_{-0.06}$ & $0.42^{+0.09}_{-0.09}$ & $5.21^{+0.10}_{-0.10}$ & $0.27^{+0.13}_{-0.13}$ & $25.67^{+0.17}_{-0.17}$ & $294.8^{+1.8}_{-1.8}$ \\
VW Cha & 1 & $93^{+7}_{-7}$ & $21^{+4}_{-5}$ & $338^{+9}_{-10}$ & $44^{+4}_{-5}$ & $9^{+4}_{-4}$ & $24^{+6}_{-6}$ & $8.7^{+2.6}_{-2.6}$ & $24^{+4}_{-4}$ & $78^{+3}_{-3}$ & $6.6^{+2.2}_{-2.2}$ & $209.2^{+2.7}_{-2.7}$ & $1232^{+51}_{-50}$ \\
VW Cha & 2 & $70^{+5}_{-5}$ & $12^{+3}_{-3}$ & $152^{+7}_{-7}$ & $32^{+3}_{-3}$ & $3.8^{+2.4}_{-2.4}$ & $15^{+6}_{-6}$ & $8^{+3}_{-3}$ & $11^{+4}_{-4}$ & $65.6^{+2.3}_{-2.3}$ & $3.1^{+1.6}_{-1.6}$ & $134.0^{+2.1}_{-2.2}$ & $890^{+60}_{-60}$ \\
VW Cha & 3 & $78^{+7}_{-7}$ & $14^{+5}_{-5}$ & $232^{+10}_{-10}$ & $43^{+5}_{-5}$ & $11^{+4}_{-4}$ & $29^{+9}_{-9}$ & $5^{+5}_{-5}$ & $22^{+6}_{-6}$ & $76^{+4}_{-4}$ & $12^{+3}_{-3}$ & $268^{+4}_{-4}$ & $1260^{+60}_{-50}$ \\
VW Cha & 4 & $76^{+11}_{-11}$ & $26^{+7}_{-7}$ & $280^{+12}_{-12}$ & $66^{+6}_{-7}$ & $14^{+6}_{-6}$ & $48^{+10}_{-10}$ & $14^{+5}_{-5}$ & $23^{+7}_{-7}$ & $95^{+5}_{-5}$ & $13^{+4}_{-4}$ & $380^{+5}_{-5}$ & $1750^{+60}_{-60}$ \\
VW Cha & 5 & $80^{+6}_{-6}$ & $13^{+4}_{-4}$ & $221^{+7}_{-7}$ & $49^{+4}_{-4}$ & $12^{+3}_{-3}$ & $26^{+7}_{-7}$ & $12^{+3}_{-3}$ & $20^{+5}_{-5}$ & $62^{+3}_{-3}$ & $4^{+3}_{-3}$ & $232^{+4}_{-4}$ & $880^{+60}_{-60}$ \\
\enddata
\label{tab:linelum}
\tablecomments{Integrated line luminosities are presented in units of $10\times10^{29} \, ergs \, s^{-1}$. Listed uncertainties are the $16^{th}$ and $84^{th}$ percentiles of the posterior, which is nearly symmetric for all lines included here. Line luminosities reported here have been corrected for extinction. The values reported here for the doublet/multiplet features are the blended line luminosities integrated over a narrow wavelength range covering the feature. }
\end{deluxetable*}
\end{rotatetable*}

\bibliography{biblio}

\end{document}